%% file: main.tex
\documentclass[10pt,a4paper]{article}
\usepackage[utf8]{inputenc}
\usepackage{geometry}
\usepackage{amsmath}
\usepackage{amssymb}
\usepackage{mathtools}
\usepackage[hidelinks]{hyperref}
\usepackage{caption}
\usepackage[font=small]{caption}

\usepackage{enumerate}
\usepackage{multicol}
\usepackage[round]{natbib}   
\usepackage{graphicx}

\usepackage{graphics}
\usepackage{lipsum}
\usepackage{adjustbox}
\usepackage{multirow}
\usepackage{tabularx}
\usepackage{soul}
\usepackage[english]{babel}
\usepackage{csquotes}
\usepackage{wrapfig}
\usepackage{float}

\usepackage{subfig}
\usepackage{fix-cm}

\PassOptionsToPackage{normalem}{ulem}
\usepackage{mdframed}    
\usepackage{ulem}
\usepackage{authblk}
\usepackage{tabularx,booktabs}
\usepackage{wrapfig, blindtext}
\newcolumntype{Y}{>{\centering\arraybackslash}X}

\usepackage{fancyhdr}
\usepackage{lastpage} 
\usepackage{tcolorbox}
\tcbuselibrary{theorems}
\makeatother
\newtcbtheorem[number within=section]{mytheo}{Theorem}%
{colback=white,colframe=black!35!black,fonttitle=\bfseries}{th}

\usepackage{xcolor}

\usepackage{dsfont}

\usepackage{titlesec} 

\titleformat{\section}{\fontsize{13}{12}\selectfont\bfseries}{\thesection}{1em}{}
\titleformat{\subsection}{\normalsize\bfseries}{\thesubsection}{1em}{}

\begin{document}

\title{Estimating breakpoints between climate states in the Cenozoic Era}
\vspace{-0.9cm}
\author[1,2]{Mikkel Bennedsen}
\author[1,2]{Eric Hillebrand}
\author[3,4]{Siem Jan Koopman}
\author[1,2]{Kathrine By Larsen}

\affil[ ]{\raggedright Authors are listed in alphabetical order }
\affil[1]{\raggedright Department of Economics and Business Economics, Aarhus University, Denmark.}
\affil[2]{\raggedright  Center for Research in Energy: Economics and Markets (CoRE), Aarhus University, Denmark.}
\affil[3]{\raggedright  Department of Econometrics, Vrije Universiteit Amsterdam, The Netherlands.}
\affil[4]{\raggedright  Tinbergen Institute, Amsterdam, The Netherlands.}
\affil[ ]{\textbf{Corresponding author:} Kathrine By Larsen (\href{mailto:kblarsen@econ.au.dk}{kblarsen@econ.au.dk}) }

\date{}








\maketitle

\vspace{-1cm}

 \begin{center}
      \textbf{Abstract}
 \end{center}
 \vspace{-0.3cm}
 \noindent This study presents a statistical time-domain approach for identifying transitions between climate states, referred to as breakpoints, using well-established econometric tools. We analyze a 67.1 million year record of the oxygen isotope ratio $\delta^{18}$O derived from benthic foraminifera. The dataset is presented in \cite{Westerhold2020}, where the authors use recurrence analysis to identify six climate states. Fixing the number of breakpoints to five, our procedure results in breakpoint estimates that closely align with those identified by \cite{Westerhold2020}. By treating the number of breakpoints as a parameter to be estimated, we provide the statistical justification for more than five breakpoints in the time series. Further, our approach offers the advantage of constructing confidence intervals for the breakpoints, and it allows for testing the number of breakpoints present in the time series.

\section{Introduction}  

\cite{Westerhold2020} present a time series dataset spanning from 67.1 million years ago (Ma) to the present time, covering the Cenozoic Era. 
Using recurrence analysis, the authors identify four major climate states $-$ Hothouse, Warmhouse, Coolhouse, and Icehouse $-$ which are further divided into six states. We refer to these as \textit{Warmhouse I} (66-56 Ma), \textit{Hothouse} (56-47 Ma), \textit{Warmhouse II} (47-34 Ma), \textit{Coolhouse I} (34-13.9 Ma), \textit{Coolhouse II} (13.9-3.3 Ma), and \textit{Icehouse} (3.3 Ma - present). The climate states in the Cenozoic Era range from very warm climates to the glaciation of Earth's polar regions \citep{Zachos2001}. 
The climatic transitions contain important information about variations in Earth's climate system; see \cite{Tierney2020_past_inform_future} for a review. Our study presents a statistical approach for identifying the transitions between climate states, referred to as \emph{breakpoints}, using econometric time-domain tools proposed by \cite{BaiPerrson1998, BaiPerron2003}. This approach offers the advantages of constructing confidence intervals for the dates of the breakpoints, providing a measure of estimation uncertainty, as well as testing for the number of breakpoints in the time series.

The methodology used by \cite{Westerhold2020} to identify breakpoints is recurrence analysis, as described by \cite{Marwan2007}. 
To conduct this analysis, \cite{Westerhold2020} resampled their data at a frequency of 5 thousand years (kyr) and used both un-detrended and detrended versions of the data. This analysis leads to the identification of the six climate states.

The use of recurrence analysis to identify climate states is common in the study of paleoclimate time series, see the review in \cite{Marwan2021}. Many extensions of this methodology have been suggested to address different issues.  \cite{Goswami2018AbruptTransitions} propose a breakpoint detection method using a probability density function sequence representation of the time series, which allows for uncertainties in the time stamping of the time series.  \cite{Bagniewski2021} combine recurrence analysis with Kolmogorov–Smirnov tests to detect abrupt transitions in a time series. \cite{Rousseau2023} applies this method on the \cite{Westerhold2020} data and find similar climate states as the ones reported in \cite{Westerhold2020}. As discussed by \cite{Marwan2021}, there are several other approaches to identify transitions in paleoclimate time series. Among these, \cite{Livina_2010} developed a new statistical method of potential analysis and applied it to detect the number of states in a geophysical time series.

Our approach contributes to the existing breakpoint detection methods in paleoclimate research by applying well-established econometric tools in the time-domain, developed in \cite{BaiPerrson1998, BaiPerron2003}, to identify climate states in the paleo record. It enables the estimation of multiple breakpoints along with confidence intervals and provides procedures to estimate the number of breakpoints.

The estimation methodology of \cite{BaiPerrson1998, BaiPerron2003} necessitates a constant observation frequency and a predetermined model specification. To obtain a constant observation frequency, we use mean binning, which entails dividing the data into intervals of fixed length and calculating the mean in each bin. We explore three different model specifications and implement these using the R-package by \cite{R-mbreaks}. The first model is a state-dependent mean model, which posits an abrupt break in the mean of $\delta^{18}$O for each climate state. The second model generalizes this by including a state-independent autoregressive term, which makes the transitions between states more gradual. The final model extends the second model by letting the autoregressive term be state-dependent as well, allowing for state-specific transition dynamics. All models incorporate an error term with state-dependent variance. Given that the time series appears state-wise non-stationary, meaning that the mean and/or
the variance of the time series vary over time within a state, we conduct a simulation study to demonstrate the applicability of the approach by \cite{BaiPerrson1998, BaiPerron2003} in this non-stationary setting.

The number of breakpoints is a parameter in the statistical approach of \cite{BaiPerrson1998, BaiPerron2003}. Fixing the number of breakpoints to five, the resulting breakpoint estimates align closely with those identified by \cite{Westerhold2020} across various binning frequencies and model specifications. This demonstrates the robustness of the approach and corroborates the dating of the climate states of \cite{Westerhold2020} with statistical analysis in the time domain. However, when we allow the number of breakpoints to vary and treat this as a parameter to be estimated  using information criteria, we find strong statistical evidence for the presence of more than five breakpoints. Specifically, our analysis suggests that Warmhouse II and Coolhouse II can each be split into two separate substates.

The remainder of the paper is structured as follows: 
In Section \ref{Data}, we present the $\delta^{18}$O dataset and climate states by \cite{Westerhold2020}. In Section \ref{Methodology}, the statistical breakpoint detection methodology applied in this paper is outlined. In Section \ref{Analysis and Results}, we conduct the analysis and discuss the results. Section \ref{Conclusion_and_outlook} concludes. The finite sample performance of the methodology under state-wise non-stationarity is investigated in a simulation study in Appendix \ref{simulation_study}.

\section{Data}

\label{Data}

\label{Westerhold_data}

 The paleoclimate variable $\delta^{18}$O measures the ratio of $^{18}$O to $^{16}$O in the shells of benthic foraminifera obtained from ocean sediment cores, relative to a standard sample.
The weight difference between the oxygen isotopes leads to an inverse relationship between $\delta^{18}$O and ocean temperatures; see for instance 
\cite{epstein1951} and \cite{shackleton1967}.

\begin{figure}[H]
    \centering
    \includegraphics[width=0.75\textwidth]{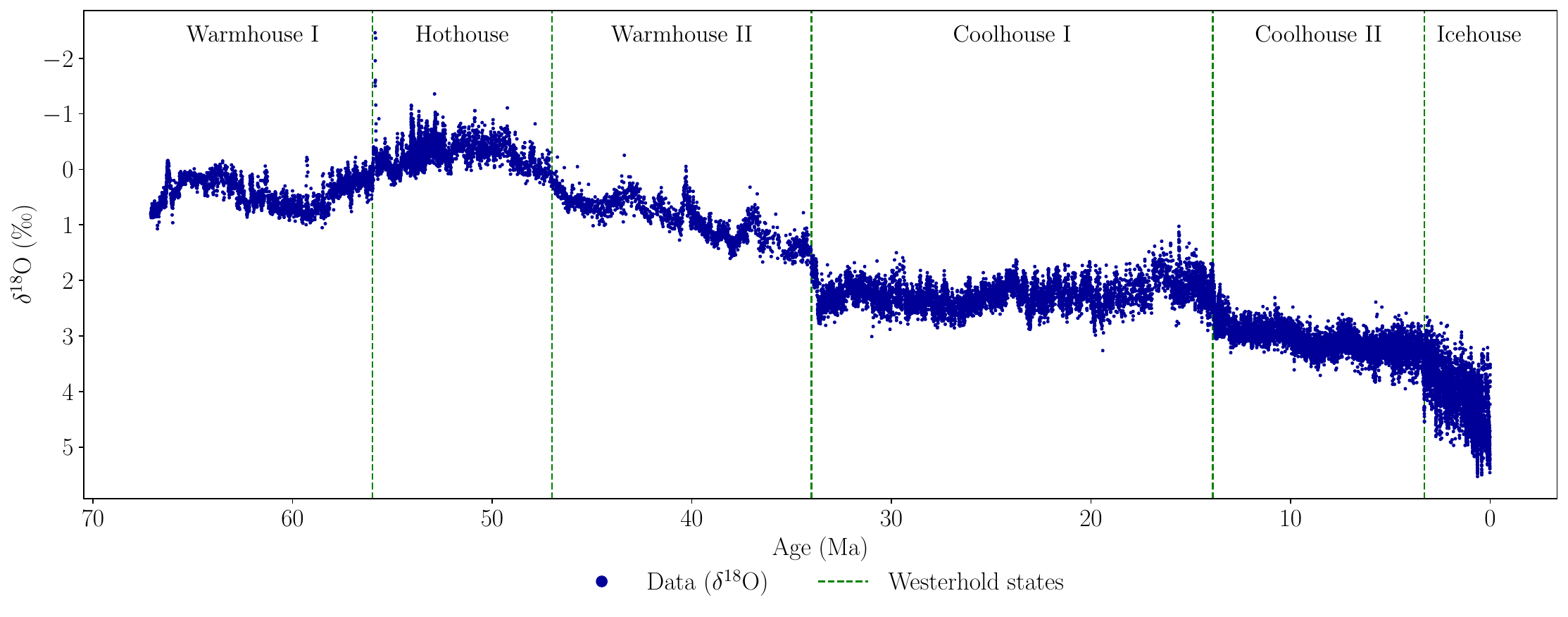}

    \includegraphics[width=0.75\textwidth]{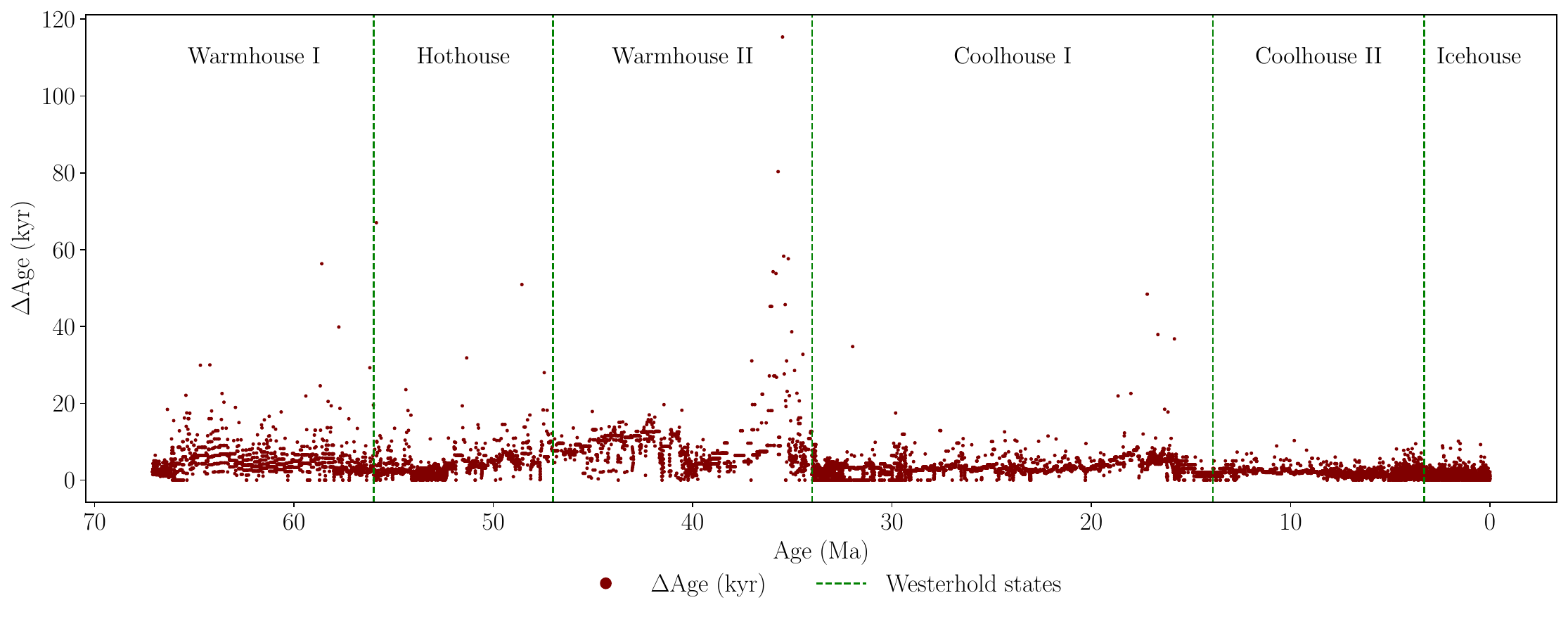}
    \vspace{-0.3cm}
    \caption{Top panel: $\delta^{18}$O data from \cite{Westerhold2020}. The order of the vertical axis is reversed, following standard practice. Bottom panel: The time between data points of $\delta^{18}$O data measured in kyr.
    The vertical dashed lines show transitions between the climate states by \cite{Westerhold2020}. The horizontal axis represents time, measured in millions of years before present.}
    \label{fig:raw_westerhold}
\end{figure}

In this paper, we use the dataset provided by \cite{Westerhold2020}, which compiles measurements of oxygen and carbon isotope ratios from benthic foraminifera across 34 different studies and 14 ocean drilling locations into a single data file.
Our study focuses on the $\delta^{18}$O record, specifically the correlation-corrected observations of $\delta^{18}$O (column “benthic d18O VPDB Corr” from the data file).
\cite{Westerhold2020} provide an estimated chronology of the data, which has accuracy ranging from $\pm 100$ kyr in the older part of the sample period to $\pm 10$ kyr in the younger part. We ignore the uncertainty of the time stamps in this study. 
 The data cover the period 67.10113 Ma to 0.000564 Ma, and we order the observations from oldest to most recent. We remove the 74 missing values in the record, leaving us with 24,259 data points. The top panel of Fig. \ref{fig:raw_westerhold} shows the $\delta^{18}$O data with the breakpoints between the climate states as identified by \cite{Westerhold2020}. Summary statistics of the dataset for the full sample length and for each climate state are presented in Appendix \ref{tab:summary_binned_state}.

The $\delta^{18}$O data present unique challenges due to its irregular nature and intermittent gaps. The bottom panel of Fig. \ref{fig:raw_westerhold} shows the time series of increments between consecutive time stamps in the record. This graph reveals that the time series is relatively sparse in the older part of the record and relatively dense in the younger part. The average time between two adjacent data points is approximately 2.8 kyr, and the longest gap between data points is approximately 115.4 kyr. There are 533 occurrences of gaps between two data points lasting longer than 10 kyr. Moreover, there are 591 instances of multiple observations at the same time stamp, with up to four simultaneous observations.

\section{Methodology}
\label{Methodology}

In this section, we present and discuss the methodology developed by \cite{BaiPerrson1998, BaiPerron2003} for estimating breakpoints in linear regression models. Section \ref{model} provides an outline of the framework for detecting multiple structural changes in linear regression coefficients. Section \ref{specification} discusses the model specifications employed in this study for breakpoint estimation. Section \ref{Implementation} outlines the implementation of the model specifications.

\subsection{General framework}
\label{model}

The method is based on minimizing the sum of squared residuals while treating the breakpoints as unknown parameters to be estimated \citep{BaiPerrson1998, BaiPerron2003}. Consider a linear regression framework for the dependent variable $y_t$, for $t=1,\hdots, T$, and with $m$ breakpoints, corresponding to $m+1$ distinct states in the sample. The general model equation is  
\begin{align}
y_t &= x_t^{\prime} \beta + z_t^{\prime} \delta_j + u_t, \quad \quad t = T_{j-1} + 1, \ldots, T_j,
\label{eq:full_model}
\end{align}
with $j = 1, \ldots, m+1$. The $m$ break dates are denoted by $\left(T_1, \ldots, T_m\right)$, with the convention that $T_0=0$ and $T_{m+1}=T$, and $u_t$ is a disturbance term with mean zero and variance $\sigma_j^2$. The  $(p\times 1)$-vector $x_t$ and the $(q \times 1)$-vector $z_t$ comprise two sets of covariate vectors, for which $\beta$ is the state-independent vector of coefficients and $\delta_j$ is the state-dependent vector of coefficients. Since only specific coefficients are subject to structural breaks, this model is referred to as a partial structural change model. Moreover, we consider
 breaks in the variance of $u_t$ at the break dates $T_1, \hdots, T_m$, such that $\sigma_i^2 \neq \sigma_j^2$ for $i\neq j$. The parameters $\beta$ and $\delta_j$ are estimated alongside the breakpoints but are not of primary interest here.

We initially treat the number of breakpoints, $m$, as known and estimate the coefficients and the breakpoints using a sample of $T$ observations of $\left\{y_t, x_t, z_t\right\}$. The estimation method is based on least squares for both the coefficients and the breakpoints. 
For each possible set of  $m$ breakpoints $\left(T_1, \ldots, T_m\right)$ denoted as $\left\{T_i\right\}_{i=1}^m$, we obtain estimates of $\beta$ and $\delta_j$ by minimizing the sum of squared residuals (SSR), that is,
\begin{align}
 SSR &= \sum_{j=1}^{m+1} \sum_{t=T_{j-1}+1}^{T_j}\left(y_t-x_t^{\prime} \beta-z_t^{\prime} \delta_j\right)^2,
   \intertext{where $\beta$ is common to all states, while $\delta_j$ is specific for the state $j$, which is the period between $T_{j-1}+1$ and $T_j$. The resulting estimated coefficients are denoted as $\hat{\beta}\left(\left\{T_i\right\}_{i=1}^m\right)$ and $\hat{\delta}\left(\left\{T_i\right\}_{i=1}^m\right)$. These coefficients are then used to determine the SSR associated  with each set of breakpoints,}
   SSR_T\left(\left\{T_i\right\}_{i=1}^m\right) &\equiv \sum_{j=1}^{m+1} \sum_{t=T_{j-1}+1}^{T_j}\left(y_t-x_t^{\prime} \hat{\beta}\left(\left\{T_i\right\}_{i=1}^m\right)-z_t^{\prime} \hat{\delta}\left(\left\{T_i\right\}_{i=1}^m\right)\right)^2.
   \intertext{ The estimated breakpoints are then given by}
     \left(\hat{T}_1, \ldots, \hat{T}_m\right)&=\underset{T_1, \ldots, T_m}{\operatorname{argmin}}\;
    SSR_T\left(\left\{T_i\right\}_{i=1}^m\right).
\end{align}
The minimization is conducted over all partitions $\left(T_1, \ldots, T_m\right)$ such that $T_j-T_{j-1} \geq \dim ( z_t)$ to ensure that there are enough data points to estimate the parameters $\delta_j$ in each partition. This procedure leads to estimated parameters for the $m$ breakpoints, i.e., $\{\hat{T}_i\}_{i=1}^m$, $\hat{\beta}=\hat{\beta}\left(\{\hat{T}_i\}_{i=1}^m\right)$, and $\hat{\delta}=\hat{\delta}\left(\{\hat{T}_i\}_{i=1}^m\right)$. Since the possible combinations of the placement of the breakpoints is finite, this optimization can be conducted using a grid search, which can be computationally heavy, especially for many breakpoints. \cite{BaiPerron2003} introduce an efficient method for determining the global minimizers. 

An essential advantage of this framework is that it allows for constructing confidence intervals for the breakpoints, something that is not available for the recurrence analysis approach implemented in \cite{Westerhold2020}. 
The construction of confidence intervals is based on the asymptotic distribution of the break dates. The convergence results for the construction of confidence intervals rely on a number of assumptions \citep[see][]{BaiPerron2003}, which may possibly be violated for the $\delta^{18}$O data. To examine whether the framework is adequate for the type of data studied here, we have conducted a large simulation study, of which the details are reported in Appendix \ref{simulation_study}. We discuss the results from the simulation study and how they relate to the analysis of the $\delta^{18}$O data in Section \ref{Analysis and Results}.

\subsection{Model specifications}
\label{specification}
    
In this section, we introduce the model specifications employed in this paper for estimating breakpoints. Three distinct specifications are considered, referred to as the ``Mean'', ``Fixed AR'', and ``AR'' models, where AR refers to the autoregressive model of order one with intercept. These are all special cases of the framework outlined in Eq. \eqref{eq:full_model}. The simplest among them, the Mean model, is specified as follows,
    \vspace{-0.1cm}
\begin{align}
  y_t &= c_j + u_t, \quad \quad t=T_{j-1}+1, \ldots, T_j,
    \label{eq:mean}
    \end{align}
    for $j=1, \hdots, m+1$, where $c_j$ is the state-dependent intercept and $u_t$ is an error term. This model is equivalent to setting $x_t=0$, $z_t=1$, and $\delta_j=c_j$ in Eq. \eqref{eq:full_model}. A breakpoint in this model specification leads to an abrupt change in the mean of the dependent variable $y_t$. 
    
    The Fixed AR model extends the Mean model by incorporating an autoregressive term. We obtain the model
    \begin{align}
    y_t &= c_j + \varphi y_{t-1}+ u_t, \quad \quad  t=T_{j-1}+1, \ldots, T_j,
       \label{eq:fixedAR}
       \end{align}
       for $j=1, \hdots, m+1$, where $y_{t-1}$ is the dependent variable lagged by one period, and $\varphi$ is the autoregressive coefficient that is constant over the whole sample. In this model, the effect of a change in the coefficient $c_j$ is more gradual since it depends on the autoregressive dynamics. The Fixed AR model is obtained from Eq. \eqref{eq:full_model} by specifying $x_t=y_{t-1}$, $\beta=\varphi$, $z_t=1$, and $\delta_j=c_j$.

      The general AR specification also allows the autoregressive term to be state-dependent, resulting in the AR model,
      \begin{align}
    y_t &= c_j + \varphi_j y_{t-1}+ u_t, \quad \quad  t=T_{j-1}+1, \ldots, T_j,
       \label{eq:AR}
\end{align}
for $j=1, \hdots, m+1$, where the autoregressive coefficient $\varphi$ in Eq. \eqref{eq:fixedAR} is now state-dependent and is denoted by $\varphi_j$.
This model is obtained from Eq. \eqref{eq:full_model} by setting $x_t=0$, $z_t=(1, y_{t-1})$, and $\delta_j=(c_j, \varphi_j)$. Here, both the intercept and the autoregressive coefficient are state-dependent. Thus, the three specifications are nested: The AR model is the most general; the Fixed AR model is nested in the AR model by setting $\varphi_1=\varphi_2= \hdots = \varphi_{m+1}$, and the Mean model is nested in the Fixed AR model by setting $\varphi=0$.

\subsection{Implementation}
\label{Implementation}

The models are implemented using the \textit{mbreaks} R-package \citep{R-mbreaks} based on the methodology of \cite{BaiPerrson1998, BaiPerron2003}. For all model specifications, we set the minimum length of a state, $h$, to 2.5 million years (Myr), facilitating the estimation of shorter climate states. 
Also, we let the variance of the error term, denoted as $\sigma^2_j$, be state-dependent.

  As outlined by \cite{BaiPerron2003}, no serial correlation is permitted in the errors of the regressions. 
  However, the time series of $\delta^{18}$O is likely subject to both autocorrelation and heteroscedasticity.
The assumption of no serial correlation in the errors may be violated in the model specifications considered in this paper, since the incorporation of only up to one lag in the covariates is unlikely to remove serial correlation in the errors.

 To address these issues, we use the autocorrelation and heteroscedasticity consistent (HAC) covariance matrix estimator with prewhitening in our implementations. The prewhitening procedure, proposed by \cite{prewhite}, entails applying an autoregressive model with one lag to $z_t \hat{u}_t$, where $\hat{u}_t$ denotes the residuals. 
The HAC covariance matrix estimator by \cite{robust} is then constructed based on the filtered series using the quadratic spectral kernel with bandwidth selected by an AR of order one approximation. This approach is used for both the Mean, Fixed AR, and AR models.

\section{Analysis and results}
\label{Analysis and Results}

This section presents the results of the breakpoint analysis of the $\delta^{18}$O record. The irregular sampling is addressed by data binning in Section \ref{Even Frequency}, where the stationarity properties of binned data are assessed. In Section \ref{Fix_to_five}, we fix the number of breakpoints to five as in \cite{Westerhold2020}. In Section \ref{Test_BP}, we treat the number of breakpoints as a parameter to be estimated.

\subsection{Constant data frequency}
\label{Even Frequency}

To conduct breakpoint estimation using the methodology of \cite{BaiPerrson1998, BaiPerron2003}, we need an equidistant time series. We use a binning approach to construct a dataset with evenly-spaced observations, which is common practice in the analysis of paleoclimate data; see for instance \cite{BOETTNER2021} or \cite{Reikard2021}. We divide the dataset into bins of fixed time intervals and compute the mean of the observations within each bin. In the case of gaps in the binned data, we use the values immediately preceding and succeeding the section with missing data to perform linear interpolation. We consider six different bin sizes, namely 5, 10, 25, 50, 75, and 100 kyr. Summary statistics for the full sample length and for each climate state identified by \cite{Westerhold2020} for all binning frequencies are provided in Appendix \ref{tab:summary_binned_state}.
Figure \ref{fig:compare_freq} illustrates the binning approach, with the top panel showing unaltered data and binned data at 5 kyr and 100 kyr frequencies, and the bottom panels zooming in on two sub-samples.

\begin{figure}[H]
    \centering
    \includegraphics[width=0.75\linewidth]{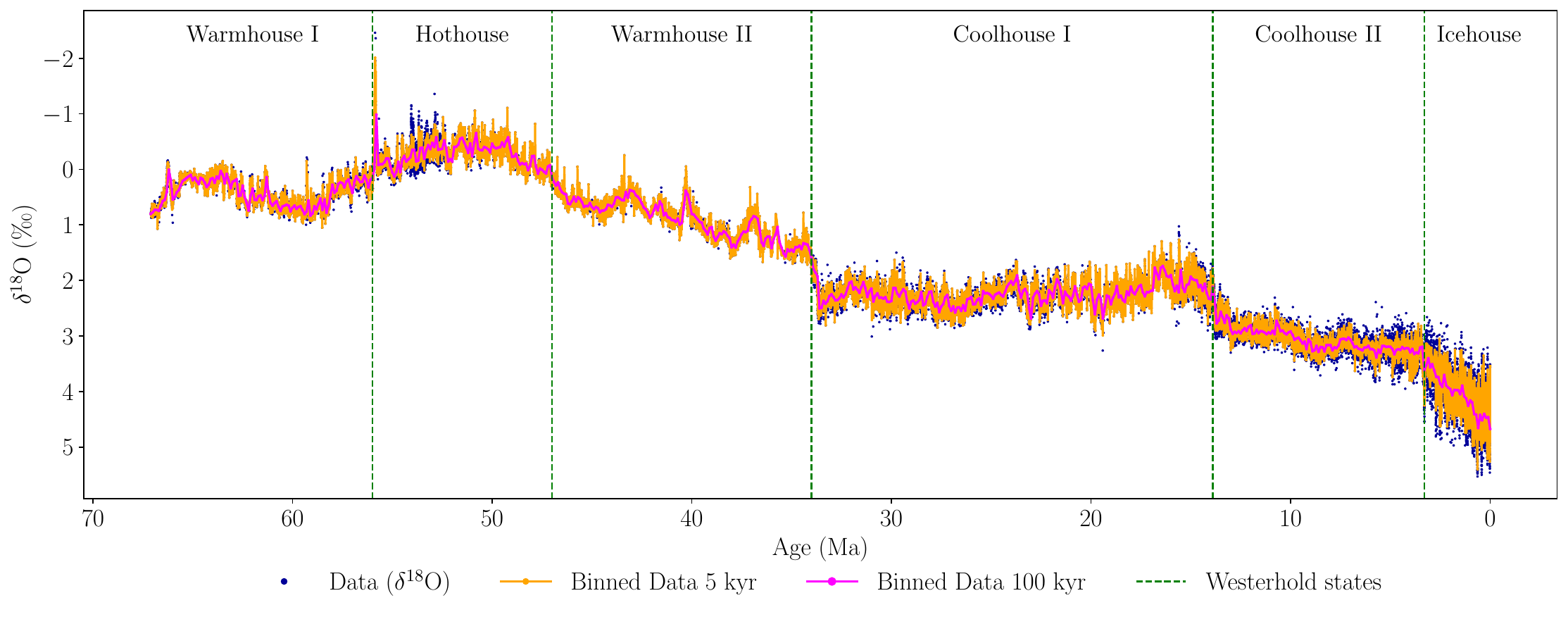}
    \vspace{0.2cm} 
    \begin{minipage}{0.5\textwidth}
        \centering
        \includegraphics[width=0.75\linewidth]{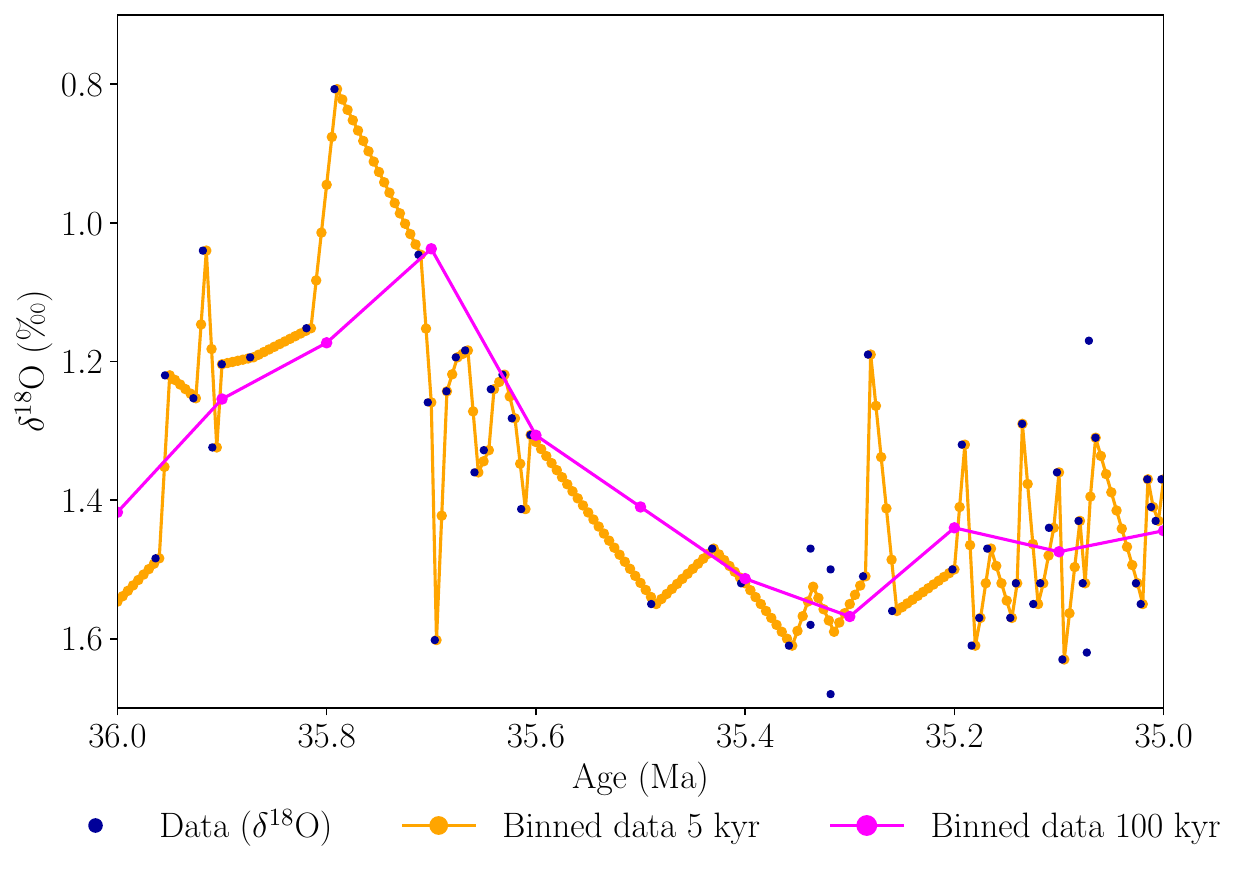}
    \end{minipage}%
    \hfill%
    \begin{minipage}{0.5\textwidth}
        \centering
        \includegraphics[width=0.75\linewidth]{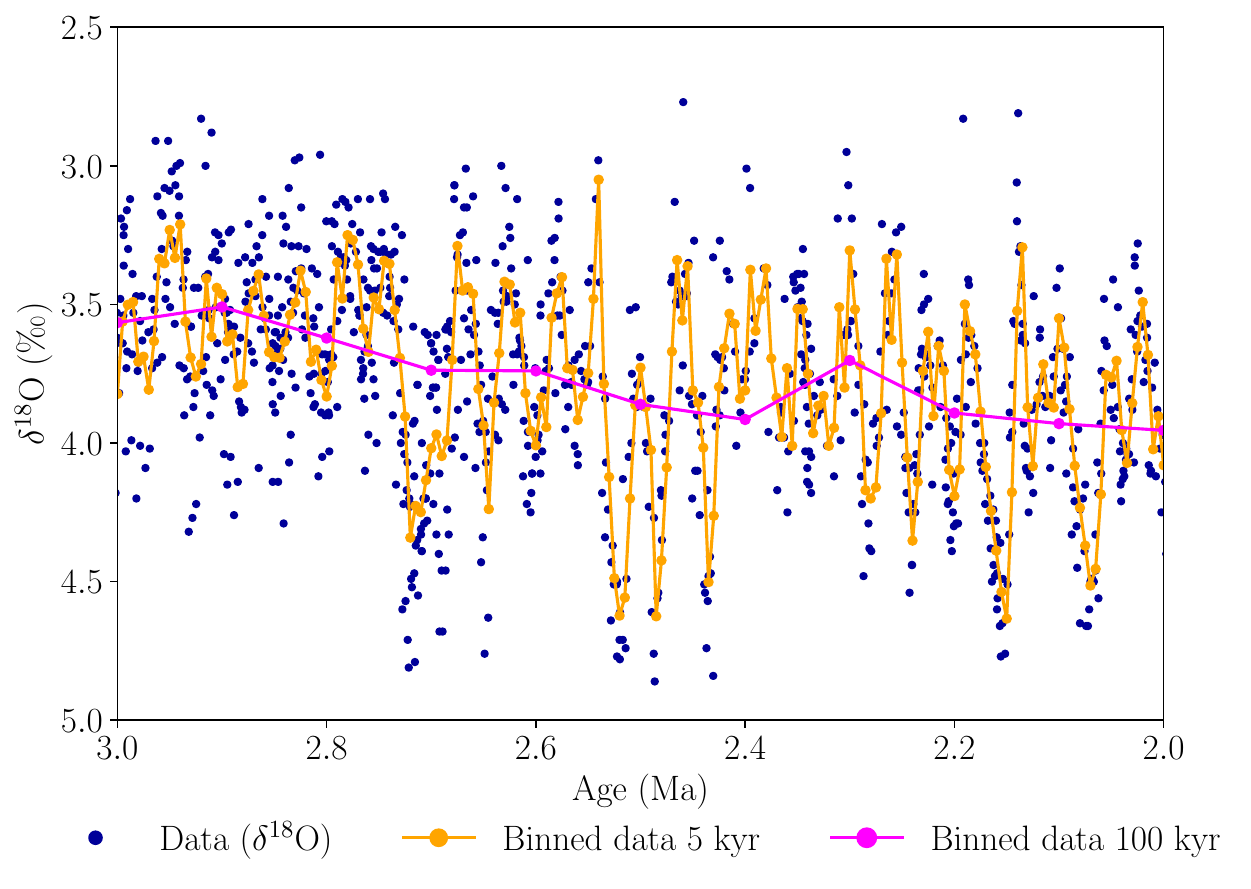}
    \end{minipage}
    
    \caption{Top panel: The original data and the 5 and 100 kyr binned data. Bottom left panel: The period 36-35 Ma. Bottom right panel: The period 3-2 Ma.}
    \label{fig:compare_freq}
\end{figure}
    \vspace{-0.3cm}

   The top panel of Fig. \ref{fig:compare_freq} shows that data binned at higher frequencies follow the variations in the dataset more closely, whereas data binned at lower frequencies tend to be smoother. The longest gap in the dataset spans approximately 115 kyr and occurs between 36 Ma and 35 Ma. The bottom panels in Fig. \ref{fig:compare_freq} zoom in on the periods 36 to 35 Ma (left) and 3 to 2 Ma (right). These plots illustrate that in case of large gaps (left), a high binning frequency results in linear interpolation between observations. This effect does not occur for periods with many observations (right), where low binning frequencies capture only a small part of the variation in the original data.
Binning offers a simple approach to handle the uneven frequency of the dataset. However, it leads to data loss at lower binning frequencies and to the introduction of artificial data points resulting from linear interpolation at higher binning frequencies. The selection of binning frequencies can therefore alter the properties of the time series, potentially misrepresenting the dynamics of the original data. 

    The theoretical framework of \cite{BaiPerrson1998, BaiPerron2003} is developed for estimating and testing for multiple breakpoints in linear regression models where the regressors are non-trending or state-wise stationary. However, the $\delta^{18}$O data appears state-wise non-stationary over most of the record.
As pointed out by \cite{Kejriwal2013}, if the time series maintains its stationarity properties over the respective states, the methods developed for stationary data are still applicable for these cases. 
However, if the process alternates between stationary and non-stationary states, the theoretical properties of the methodology are unknown. 

To investigate whether the time series is non-stationary, we apply the Augmented Dickey-Fuller (ADF) test \citep{Dickey1979}, with the null hypothesis of non-stationarity. For the entire 25 kyr binned data sample, the ADF test does not reject the null hypothesis at the 1\% significance level, indicating non-stationarity. 
However, when examining the binned data for each climate state identified by Westerhold et al. (2020) separately, the ADF test rejects the null hypothesis at the 1\% significance level for the Warmhouse II, Coolhouse I, and Icehouse states. These tests indicate the presence of state-wise non-stationarity, and we therefore need to examine whether the methodology of \cite{BaiPerrson1998, BaiPerron2003} 
is applicable to data-generating processes that are state-wise non-stationary or alternating between stationary and non-stationary states.

We conduct a simulation study to examine potential challenges in conducting breakpoint estimation on these types of data-generating processes using the three model specifications. The study is conducted for both independent and identically  distributed (i.i.d.) error terms and serially correlated error terms in Appendices \ref{simulation_studyA1} and \ref{simulation_study_serial}, respectively. 
The results show that the procedure works well with non-stationarity and is robust to processes with one stationary and one non-stationary state for Fixed AR and AR models. However, the Mean model performs poorly with highly persistent data-generating processes. 
In the case of serial correlation, the results are less conclusive, but if the states are sufficiently different, the methodology still appears effective. The study also reveals that the coverage rates of the estimated confidence intervals are generally adequate for the Fixed AR model specification in cases of large breaks. In contrast, the confidence intervals for the AR model are too narrow in many of the data-generating processes considered.

\subsection{Fixing the number of breakpoints to five}
\label{Fix_to_five}

In this section, we fix the number of breakpoints to five, which is the number found in \cite{Westerhold2020}. We estimate the breakpoints and corresponding 95\% confidence intervals for each of the binning frequencies, 5, 10, 25, 50, 75, and 100 kyr. 
In each estimation, we use a minimum state length of 2.5 Myr, allowing us to estimate relatively short climate states. The estimated breakpoints are tabulated in Appendix \ref{tab:Est_BP_5} and are shown in Fig. \ref{fig:freq_compare5}, with subfigures for the Mean, Fixed AR, and AR models. 
The estimated confidence intervals around the breakpoints are often asymmetrical. \cite{BaiPerron2003} advocate the use of asymmetric confidence intervals, as these provide better coverage rates when the data are non-stationary. 

For the Mean model presented in Fig. \ref{fig:freq_compare5}(a), it is evident that the estimated breakpoints generally remain at the same dates throughout as the binned data frequency decreases step-by-step from 5 kyr to 100 kyr. The width of the 95\% confidence intervals increases as the frequency decreases, which can be attributed to the resultant decrease in the number of binned observations available for estimation at the lower frequencies. All the breakpoints align with those identified by \cite{Westerhold2020}. A similar pattern of alignment is observed in the Fixed AR model, albeit with tighter confidence intervals, as depicted in Fig. \ref{fig:freq_compare5}(b). 

Figure \ref{fig:freq_compare5}(c) presents the findings for the AR model, which exhibits more sensitivity to the frequency of the binned data. At higher frequencies, the breakpoints tend to appear in the more recent parts of the sample. However, as the frequency decreases further, the breakpoints are estimated to be in the older parts of the sample period.
For the results using 25 kyr, we find that the estimated breakpoints from the three model specifications align closely. The estimated breakpoints align almost perfectly with those identified by \cite{Westerhold2020} and hence strongly corroborate their findings. The parameters of the three model specifications, estimated using the 25 kyr binned data with the number of breakpoints fixed at five, are provided in Appendix \ref{est_params_5}.

\begin{figure}[H]
    \centering

        \includegraphics[width=0.8\textwidth]{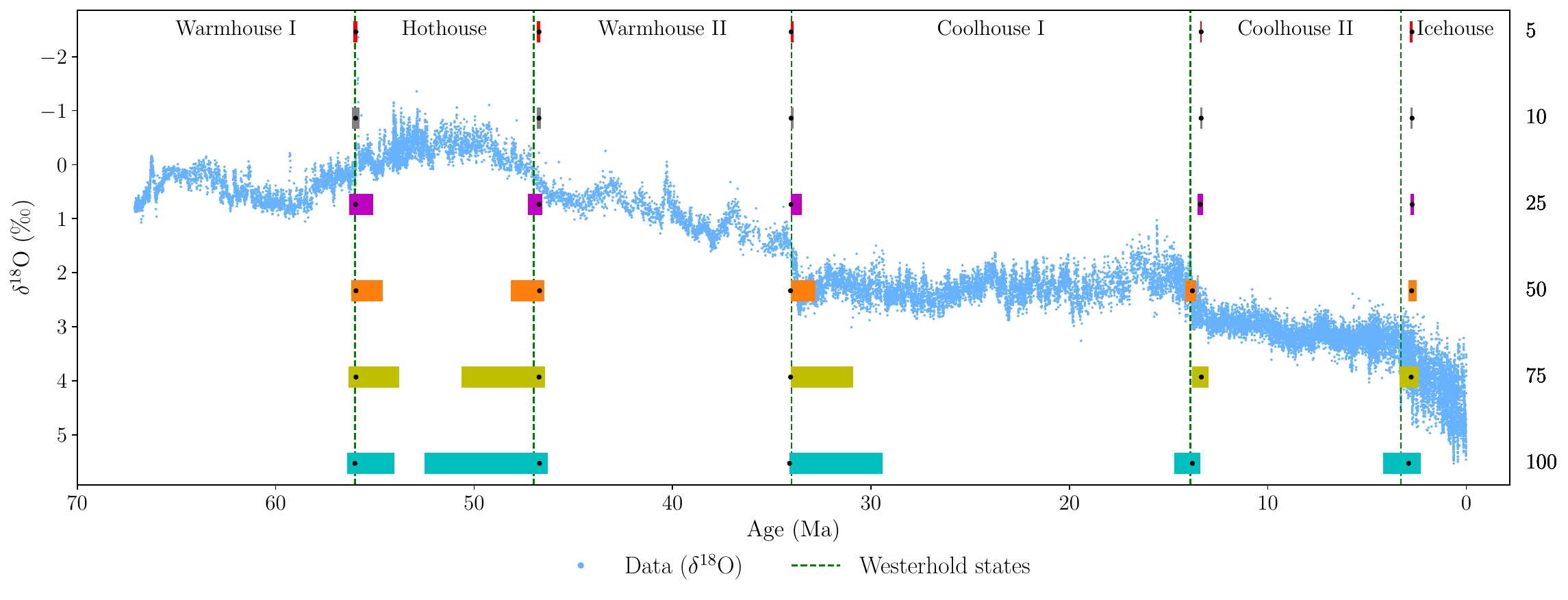}\\
        \vspace{-0.2cm}
        \text{\small (a) Mean model}\\
        \vspace{0.1cm}
        \includegraphics[width=0.8\textwidth]{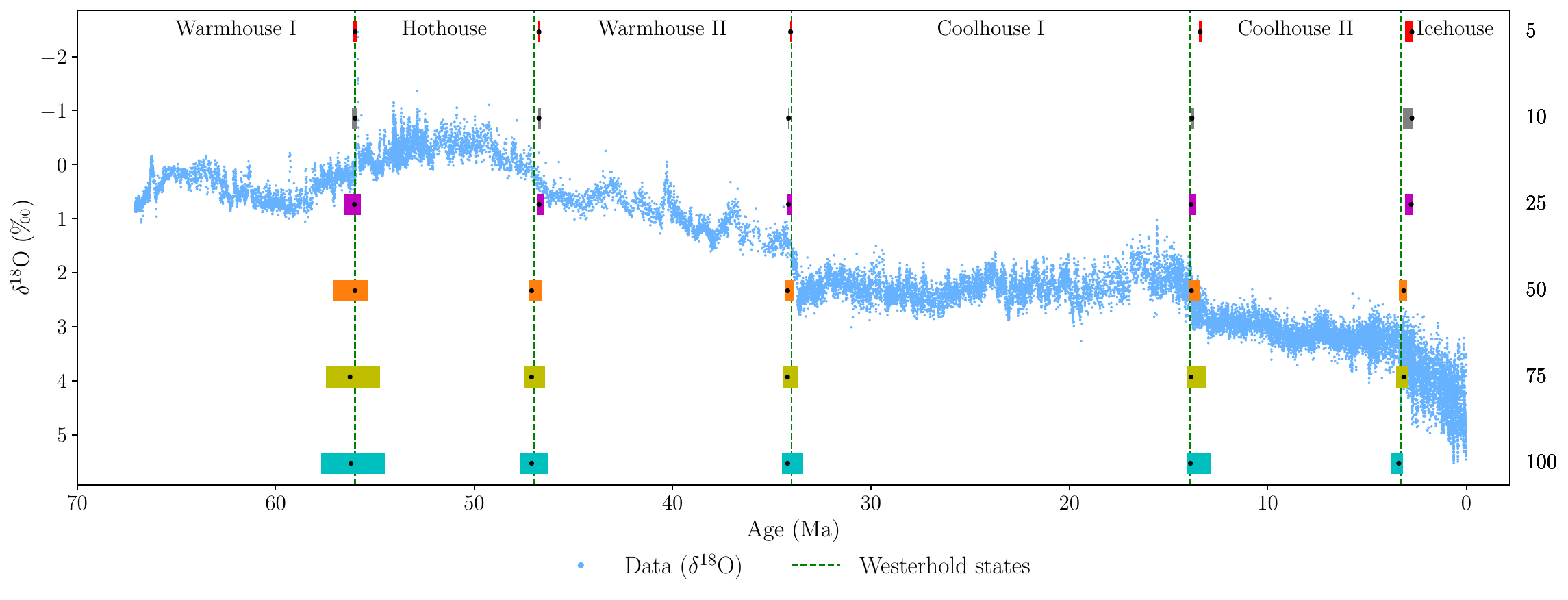}\\
         \vspace{-0.2cm}
                \text{\small(b) Fixed AR model}\\
        \vspace{0.1cm}
        \includegraphics[width=0.8\textwidth]{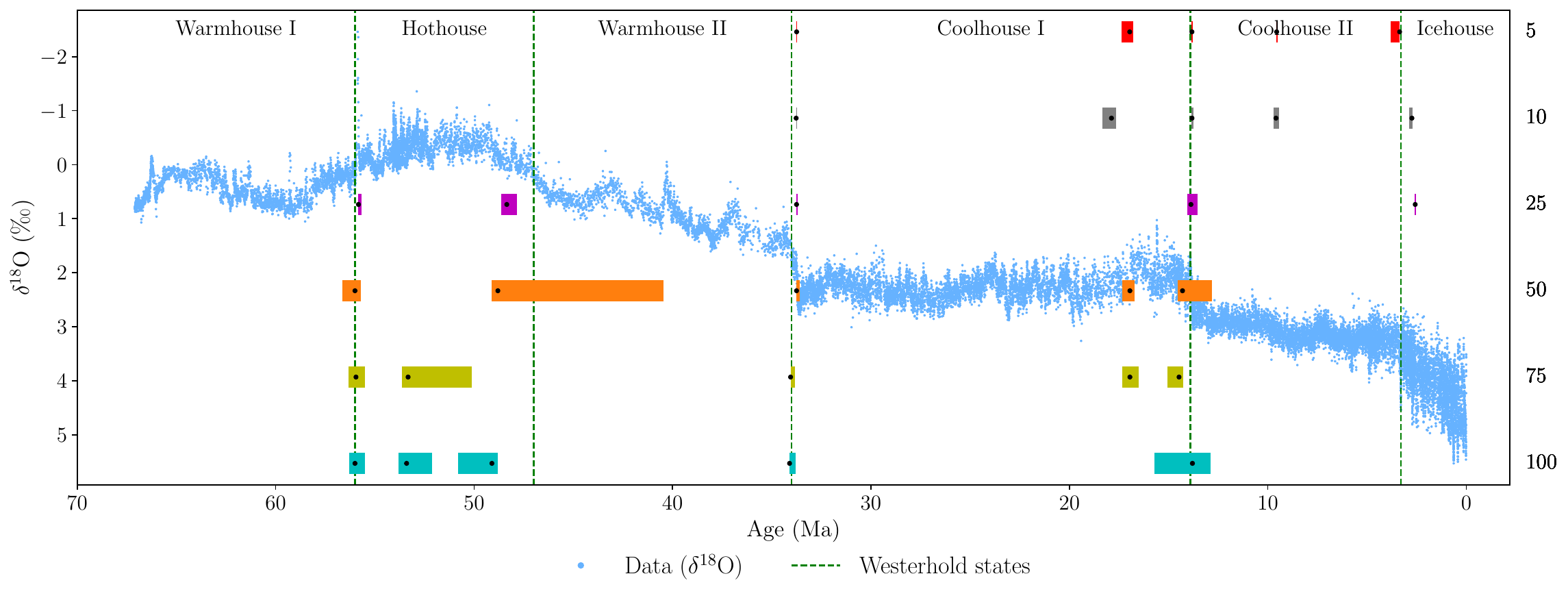}\\
         \vspace{-0.2cm}
                        \text{\small (c) AR model}\\

    \caption{A comparison of estimated breakpoints using binned data with frequencies of 5, 10, 25, 50, 75, and 100 kyr from top to bottom, fixing the number of breakpoints to five for each model specification. The black dots represent estimated breakpoints, while colored shaded rectangles indicate 95\% confidence intervals. The results overlay the $\delta^{18}$O data from \cite{Westerhold2020} and their climate states.}
    \label{fig:freq_compare5}
\end{figure}

 \vspace{-0.2cm}

As a robustness check, we re-estimate the model specifications for five breakpoints using the 25 kyr binned data reversed with respect to the time dimension, i.e., letting time run backwards. The results are shown in Appendix \ref{Reversed}. 
We find that the results of the Mean and Fixed AR models are robust to reversing the time frame, with almost unchanged estimated breakpoints. Conversely, the AR model leads to estimated breakpoints in the more recent part of the sample, resulting in breakpoints at 16.9 Ma and  9.7 Ma, which differ from those estimated using the same model and binning frequency with time running forward.

In summary, the results of the Mean and Fixed AR models exhibit robustness across different binning frequencies, whereas the AR model appears more sensitive to variations in binning frequency and the reversal of the time dimension.
As detailed in the simulation study in Appendix \ref{simulation_study}, the Mean model fails to accurately detect breakpoints in highly persistent data-generating processes.
Consequently, in what follows, we focus on the Fixed AR model for the estimation of breakpoints in the $\delta^{18}$O time series. Furthermore, we recommend using binning frequencies 10 and 25 kyr as they result in the most consistent outcomes.

\subsection{Estimating the number of breakpoints}
\label{Test_BP}

We use information criteria to estimate the number of breakpoints. We initially consider the following three criteria: the Bayesian Information Criterion (BIC) by \cite{BIC_info}, the modified Schwarz Information Criterion (LWZ) by \cite{LWZ_info}, and the modified BIC (KT) by \cite{KT_info}. For all criteria, the estimated number of breakpoints is determined as the number of breakpoints that minimizes the information criterion in question.   

\cite{bai_perron_2006} note that the BIC and LWZ criteria perform well in absence of serial correlation, but both of them lead to overestimation of the number of breakpoints in case of serial correlation in the error term.  In simulation studies, reported in Appendices \ref{simulation_studyA1} and \ref{simulation_study_serial}, we find that the KT information criterion performs poorly, and hence, we exclude it from the subsequent analysis. 
We also find that the number of breakpoints estimated using the Mean model specification is generally overestimated when employing the information criteria. For the Fixed AR and AR models, the BIC and LWZ criteria typically perform well, especially in data-generating processes with a large break. With serial correlation in the error term, the BIC criterion tends to overestimate the number of breakpoints, whereas the LWZ criterion generally performs well in the Fixed AR and AR model specifications.  

\begin{table}[H]
\fontsize{10}{11}\selectfont
  \centering
  \begin{tabular}{lcccccc}
    \hline
Bin size & \multicolumn{2}{c}{\uline{\quad\quad Mean \quad\quad}} & \multicolumn{2}{c}{ \uline{\quad\; Fixed AR \quad} } & \multicolumn{2}{c}{\uline{\quad \quad AR \quad\quad\;}} \\
    & BIC & LWZ  & BIC & LWZ & BIC & LWZ  \\
  \hline
    5  &  19  &  17  &  17 & 7  &  15 &  5   \\
    10 &  17  &  17  &  14 &  7 & 14  &   3  \\
    25 &  17  &  14  & 12  &  6 &   8 &  3   \\
    50 &  17  &  14  & 10  &  0 &  7  &   0  \\
    75 &  17  &  14  &   6 &  0 &  5  &   0  \\
    100&  17  &  12  &   6 &  0 &  5  &   0  \\
  \hline
  \end{tabular}
  \caption{ The number of breakpoints estimated using BIC and LWZ criterion for all models and binning frequencies considered. The minimum state length is set to $h=2.5$ Myr and the maximum number of breakpoints is 26.}
  \label{tab:fullINFO}
\end{table}

We use the BIC and LWZ information criteria for each model specification and binning frequency, and set the minimum state length to $h=2.5$ Myr. Table \ref{tab:fullINFO} shows the estimated number of breakpoints. There is a tendency towards higher numbers with increasing binning frequency, and for the BIC to indicate higher numbers than the LWZ criterion. For our preferred specification, the Fixed AR model with 25 kyr binning frequency, the LWZ and BIC criteria suggests six and twelve breakpoints, respectively. For a 10 kyr binning frequency, the estimated number of breakpoints are seven and fourteen, respectively. Thus, the information criteria indicate that the number of distinct climate states in the $\delta^{18}$O record is larger than the five suggested in \cite{Westerhold2020}.

\begin{figure}[H]
    \centering
        \includegraphics[width=0.82\textwidth]{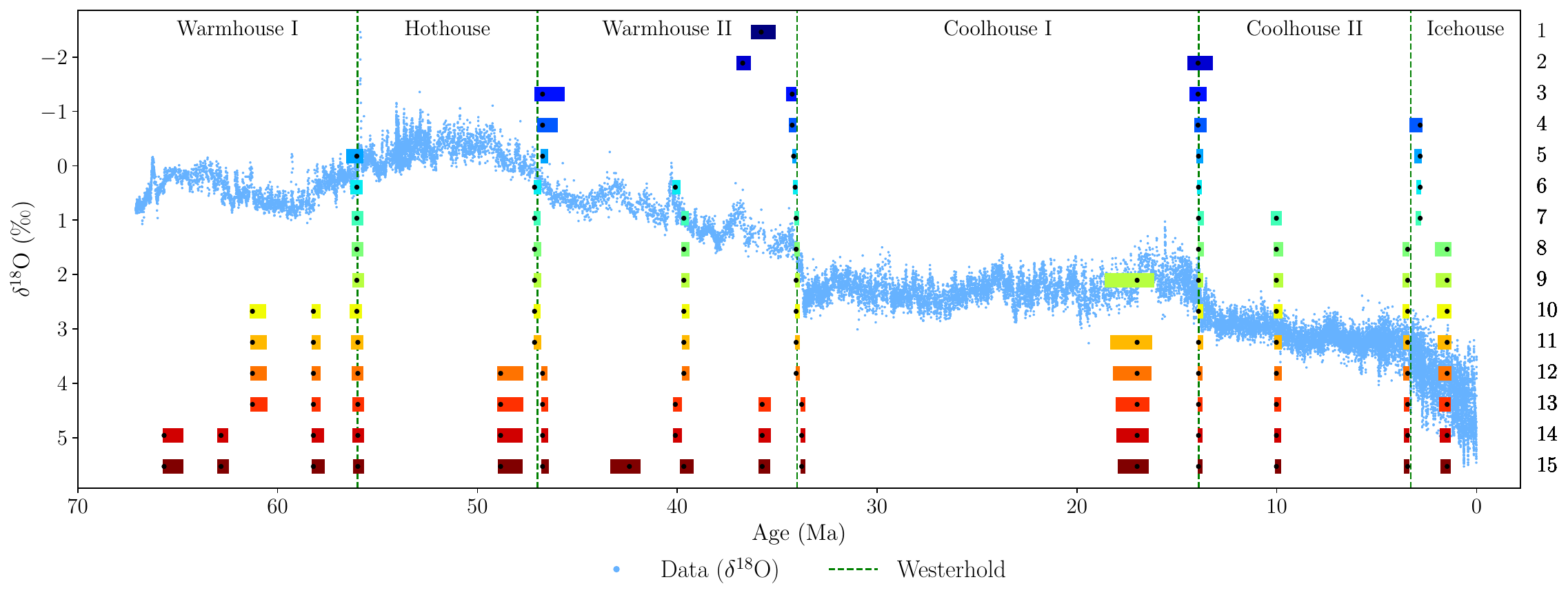}
\vspace{-0.2cm}
    \caption{
A comparison of estimated breakpoints using the Fixed AR model for one to fifteen breakpoints on 25 kyr binned data. The minimum state length is set to $h=1$ Myr. The black dots represent estimated breakpoints, while colored shaded rectangles indicate 95\% confidence intervals. The results overlay the $\delta^{18}$O data from \cite{Westerhold2020} and their climate states. }
    \label{fig:ManyBPs}
\end{figure}

To further investigate a higher number of breakpoints, we consider the estimation of up to fifteen breakpoints. 
As previously discussed, we advocate for using the Fixed AR model and mid-range binning frequency. Therefore, we estimate one to fifteen breakpoints using the Fixed AR model and 25 kyr binned data. The results are shown in Fig. \ref{fig:ManyBPs}. The same analysis conducted using the 10 kyr binned data led to nearly identical breakpoint estimates and is therefore omitted for conciseness. 

When five or more breakpoints are allowed, we find that the five  breakpoints identified by \cite{Westerhold2020} consistently coincide with five of the estimated breakpoints using our approach, indicating  the robustness of these five breakpoints. Allowing for six breakpoints, we get an additional breakpoint in Warmhouse II that remains for higher numbers of breakpoints. This breakpoint aligns with the Middle Eocene Climatic Optimum, a known climatic event. Allowing for seven, we get another one in Coolhouse II; allowing for eight, another in Icehouse; allowing for nine, another in Coolhouse I; and so on. Some of these breakpoints coincide with other climatic events, for instance, the Latest Danian Event at 62.2 Ma and the onset of the Miocene Climatic Optimum at 16.9 Ma \citep{Westerhold2020}.

The estimation results based on information criteria justify dividing the climate states Warmhouse II and Coolhouse II into two substates each at approximately 39.7 Ma and 10 Ma, respectively. This is supported by the presence of breakpoints estimated approximately at these time stamps in the estimations with seven or more breakpoints. 
Comparable findings are presented in Appendices \ref{1_15_Mean} and \ref{1_15_AR}, which detail the results of estimating one to fifteen breakpoints using the Mean and AR models, respectively, with 25 kyr binned data. 


\section{Conclusion}
\label{Conclusion_and_outlook}

This study presented a statistical time-domain approach to identify breakpoints between
climate states in the Cenozoic Era, using the econometric tools developed by \cite{BaiPerrson1998, BaiPerron2003}. We analyzed the time series of $\delta^{18}$O provided by \cite{Westerhold2020}, which we made equidistant through mean-binning. \cite{Westerhold2020} identified five breakpoints using recurrence analysis, and our analysis strongly corroborated the placements of these breakpoints across various model specifications and binning frequencies.
 Our approach offered the advantage of constructing confidence intervals for the dates
of the breakpoints, providing a measure of estimation uncertainty.
Based on the results of our simulation study, we advocate using the model specification with a state-independent autoregressive term and state-dependent intercept.

By treating the number of breakpoints as a parameter to be estimated using information criteria, we provided statistical justification for more than five breakpoints in the time series. We found evidence that Warmhouse II and Coolhouse II could each be split into two substates, which were preserved along with the five breakpoints also found by \cite{Westerhold2020} when considering seven or more breakpoints.

In concurrent research \citep{Bennedsen2024Estimating}, we address some other key challenges specific to paleo time series, including uneven time stamps, multiple observations at the same time point, and measurement errors, while taking the breakpoints identified by \cite{Westerhold2020} as given. Future research can then focus on conducting breakpoint detection using the $\delta^{18}$O time series without applying data aggregation techniques.







\par\vspace{1cm}
\noindent\textit{Code and data availability.} The data used in this study are available as the supplementary material of \cite{Westerhold2020}. The code used to conduct the analysis is based on the R-package \textit{mbreaks} by \cite{R-mbreaks} and the implementation is available upon request. 

\par\vspace{0.6cm}
\noindent\textit{Author contributions.} All authors contributed equally and are listed alphabetically.
\par\vspace{0.6cm}
\noindent\textit{Competing interests.} The authors declare that they have no conflict of interest.
\par\vspace{0.6cm}
\noindent\textit{Acknowledgements.} For helpful comments and suggestions, we thank participants at the conferences on Econometric Models of Climate Change (EMCC-VII and VIII) in Amsterdam in 2023 and in Cambridge, UK, in 2024, at the General Assembly of the EGU in Vienna in 2024, and seminar participants at Aarhus University. MB acknowledges funding from the International Research Fund Denmark under grant 7015-00018B.

\par\vspace{0.4cm}


\bibliography{references}

\appendix
\input{Appendix}

\end{document}

%% file: Appendix.tex
\section{Simulation study}
\label{simulation_study}

\subsection{Serially uncorrelated error term}
\label{simulation_studyA1}

In this appendix, we assess whether the methodology by \cite{BaiPerrson1998, BaiPerron2003} can be used to accurately estimate the number and timing of breakpoints in a state-wise non-stationary time series. We conduct 1000 simulations for each data-generating process (DGP) with a sample size of 500. All the DGPs considered have the following form,
\begin{align}
    y_t &= c_1 + \varphi_1 y_{t-1}+ \varepsilon_t, \quad \varepsilon_t \overset{\textit{i.i.d.}}{\sim} \mathcal{N} \left(0, \sigma^2\right) \quad \text{ for } t\leq T/2 \nonumber\\
    y_t &= c_2 + \varphi_2 y_{t-1}+ \varepsilon_t, \quad \varepsilon_t \overset{\textit{i.i.d.}}{\sim} \mathcal{N} \left(0, \sigma^2\right) \quad \text{ for } t> T/2.
    \label{sim_model}
\end{align}
Hence, we consider a single breakpoint in the middle of the sample interval, namely at $t=250$.
 We examine eight DGPs, each specified and described in Table \ref{table:sim_specification}.

\begin{table}[ht]
\fontsize{9}{10}\selectfont
\centering
\begin{tabular}{ccccccl}
\hline
DGP & $\sigma$ & $c_1$ & $c_2$ & $\varphi_1$ & $\varphi_2$ & Description\\
\hline
1 & 1    & 0.1 & 0.2 & 1    & 1    & Small break in the drift term of a RW\\
2 & 1    & 0.1 & 1   & 1    & 1    & Large break in the drift term of a RW\\
3 & 1    & 0.1 & 1   & 0.95 & 0.95 & Large break in the intercept and a fixed AR-coefficient\\
4 & 1    & 0.1 & 1   & 0.95 & 1    & Break in the intercept and small break in the AR-coefficient\\
5 & 1    & 0.1 & 1   & 0.5  & 1    & Break in the intercept and large break in the AR-coefficient\\
6 & 1    & 1   & 1   & 1    & 1    & RW with a drift \textit{without} a breakpoint \\
7 & 0.5  & 0.1 & 1   & 1    & 1    & Large break in the drift of a RW with low variance \\
8 & 1    & 0.1 & 1   & 0.5 & 0.5 & Large break in the intercept and a low fixed AR-coefficient\\
\hline
\end{tabular}
\vspace{-0.2cm}
 \caption{ Data-generating  processes for the simulation study and short descriptions. RW: random walk.}
\label{table:sim_specification}
\end{table}
\vspace{-0.3cm}

\begin{figure}[H]
    \centering
    \begin{minipage}{0.38\textwidth}
        \centering
        \includegraphics[width=\textwidth]{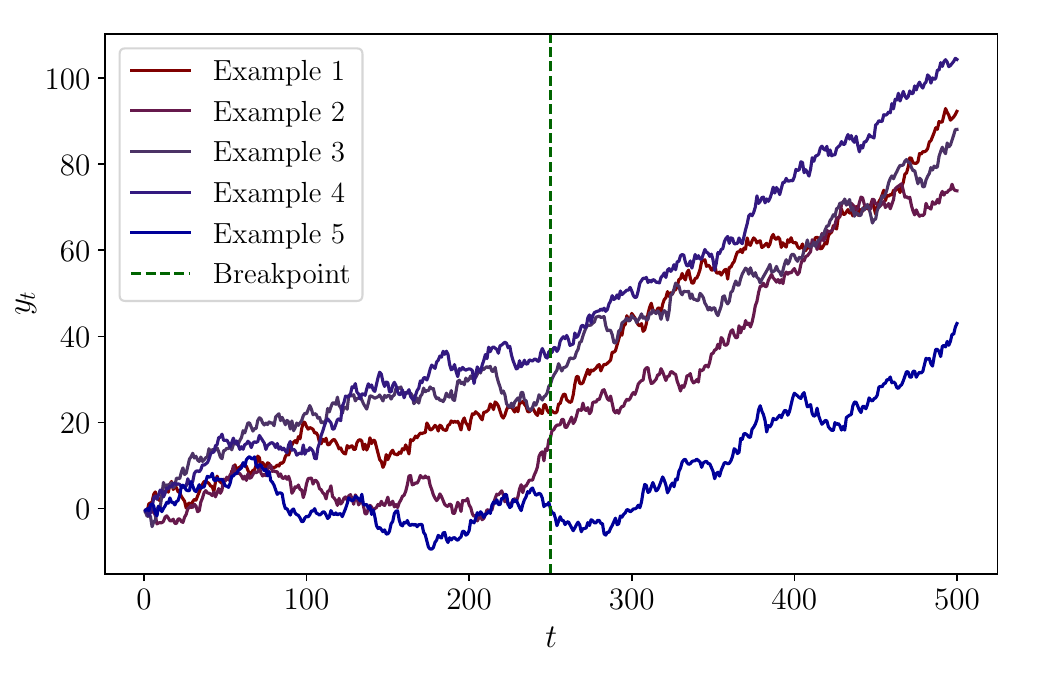}
    \end{minipage}%
    \hspace{0.5cm} 
    \begin{minipage}{0.38\textwidth}
        \centering
        \includegraphics[width=\textwidth]{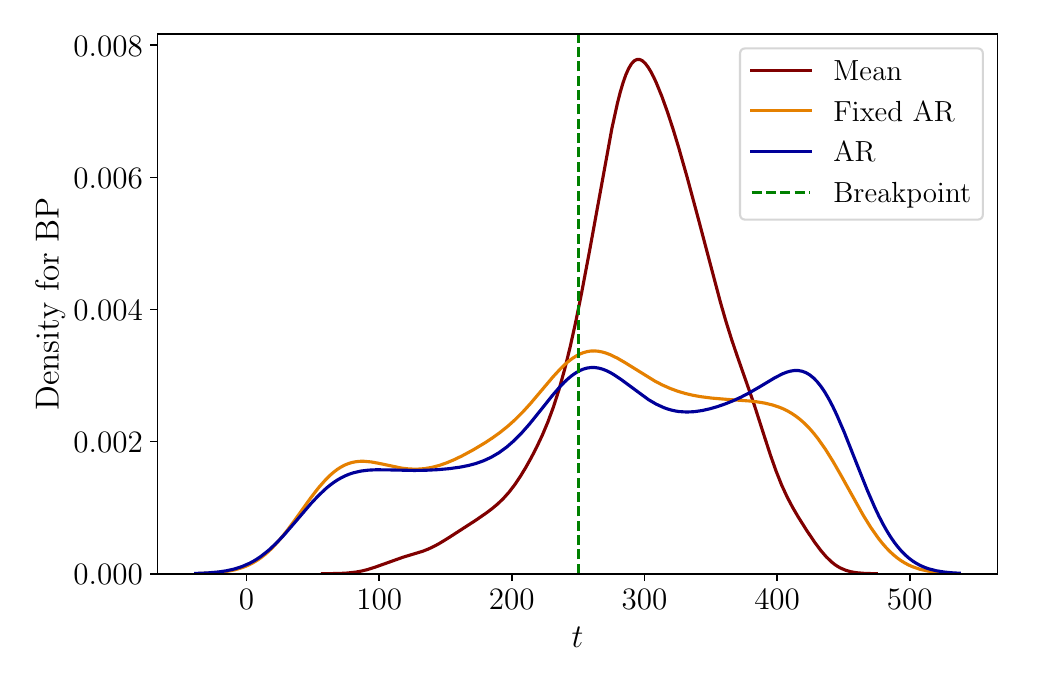}
    \end{minipage}
    
\vspace{-0.2cm}
    \caption{DGP 1: Left: Five process realizations. Right: The densities of the estimated breakpoints for each specification.}
    \label{fig:Sim_1}
\end{figure}

\begin{figure}[H]
    \centering
    \begin{minipage}{0.38\textwidth}
        \centering
        \includegraphics[width=\textwidth]{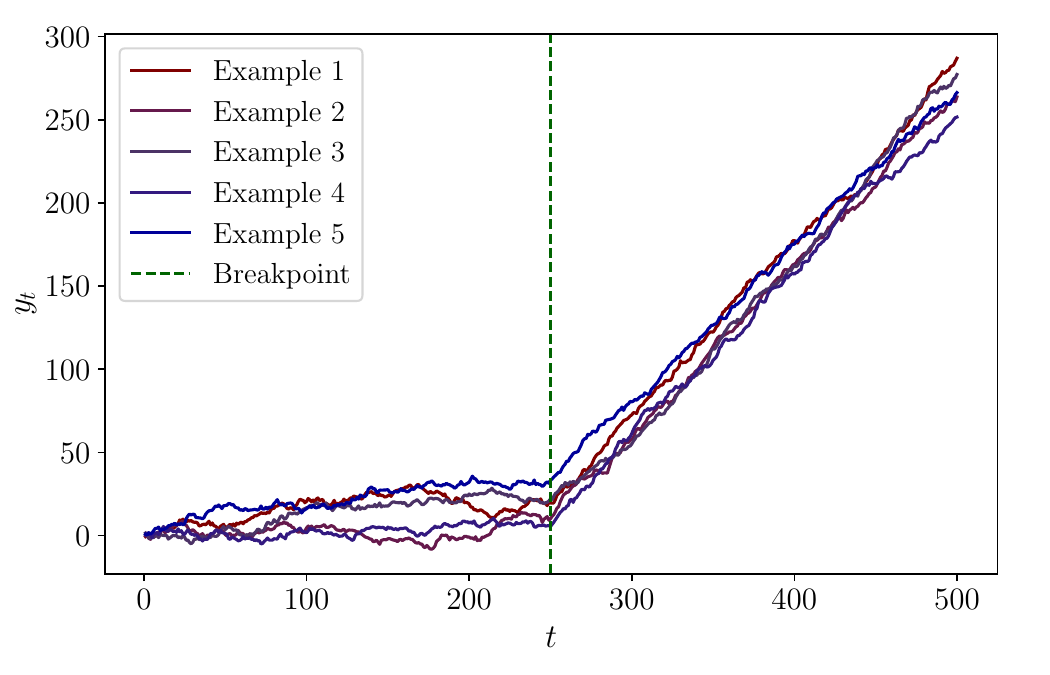}
    \end{minipage}%
    \hspace{0.5cm} 
    \begin{minipage}{0.38\textwidth}
        \centering
        \includegraphics[width=\textwidth]{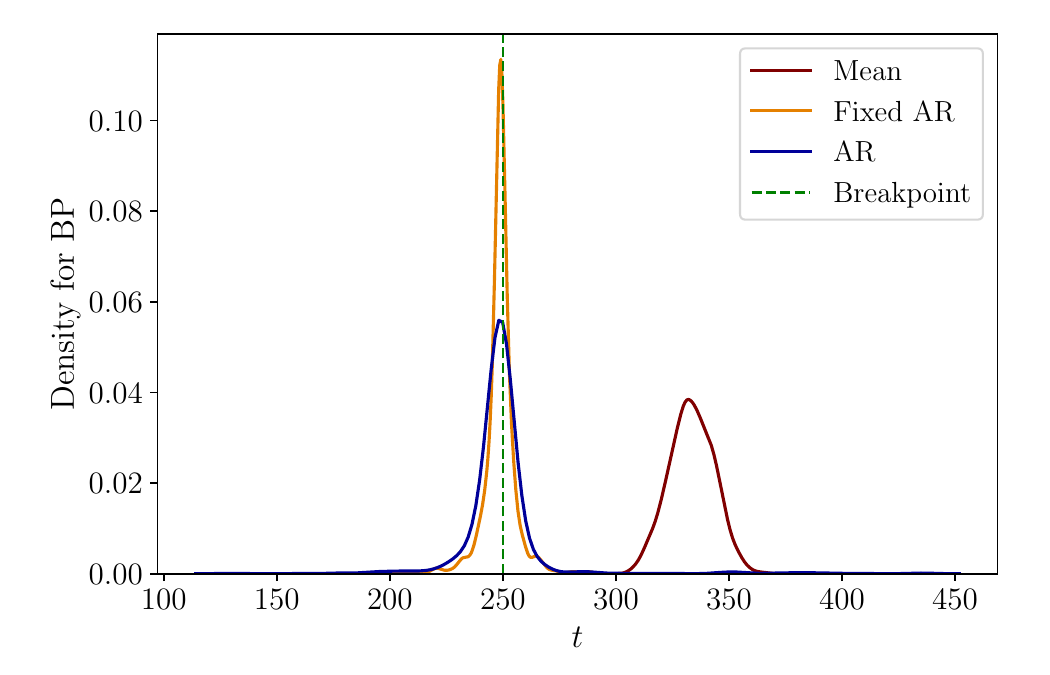}
    \end{minipage}
    \vspace{-0.2cm}
    \caption{DGP 2: Left: Five process realizations. Right: The densities of the estimated breakpoints for each specification.}
    \label{fig:Sim_2}
\end{figure}

\begin{figure}[H]
    \centering
    \begin{minipage}{0.38\textwidth}
        \centering
        \includegraphics[width=\textwidth]{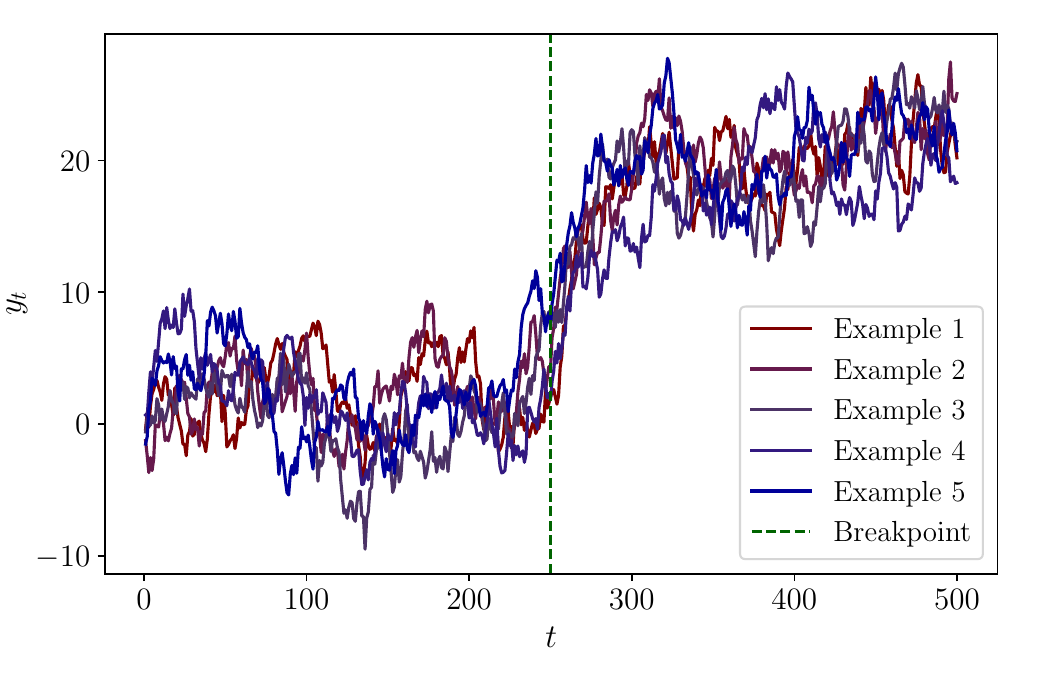}
    \end{minipage}%
    \hspace{0.5cm} 
    \begin{minipage}{0.38\textwidth}
        \centering
        \includegraphics[width=\textwidth]{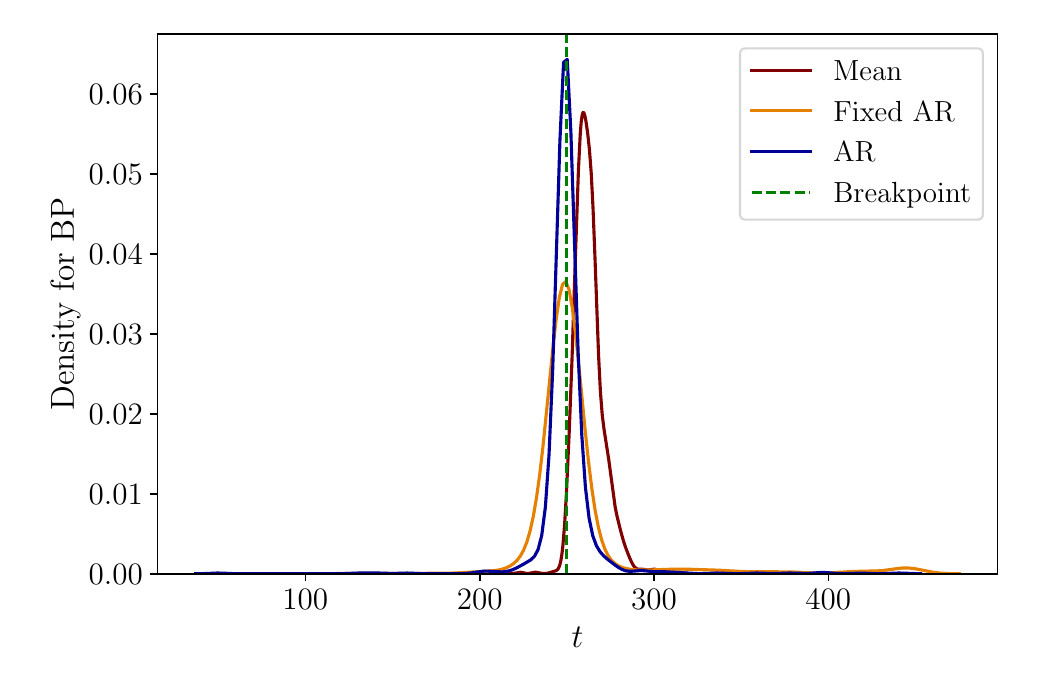}
    \end{minipage}
    \vspace{-0.2cm}
    \caption{DGP 3: Left: Five process realizations. Right: The densities of the estimated breakpoints for each specification.}
    \label{fig:Sim_3}
\end{figure}

\begin{figure}[H]
    \centering
    \begin{minipage}{0.38\textwidth}
        \centering
        \includegraphics[width=\textwidth]{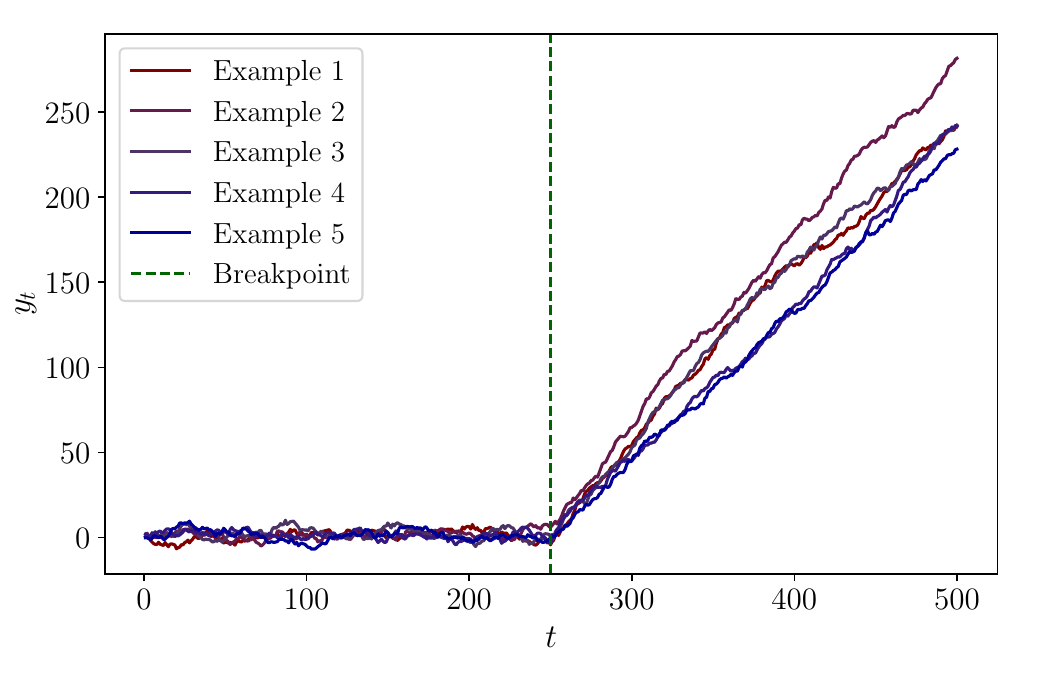}
    \end{minipage}%
    \hspace{0.5cm} 
    \begin{minipage}{0.38\textwidth}
        \centering
        \includegraphics[width=\textwidth]{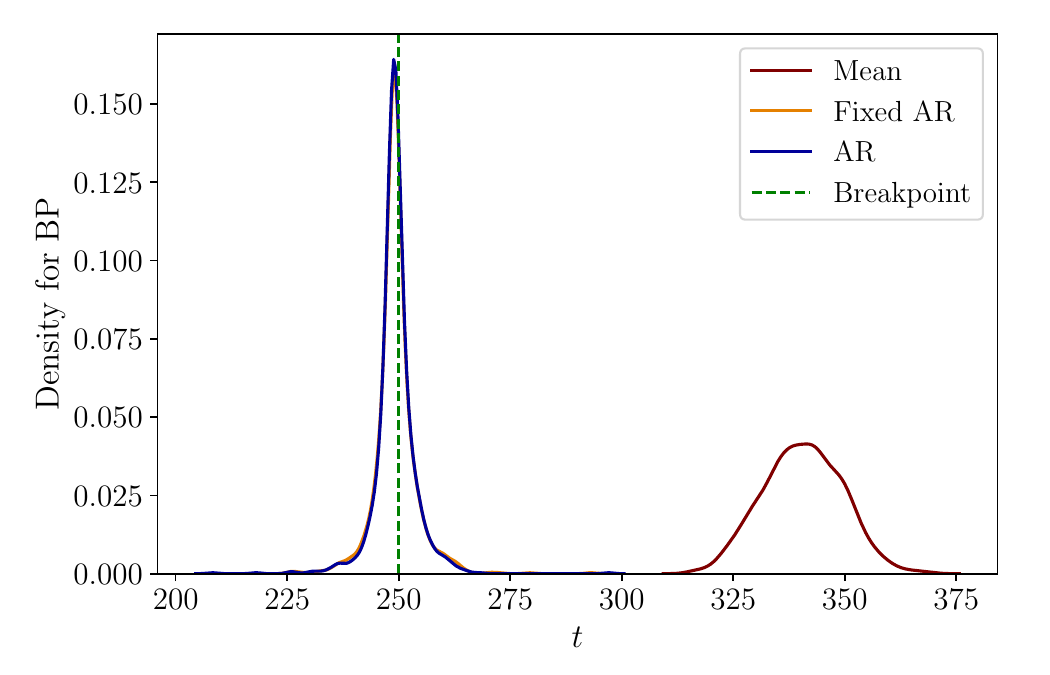}
    \end{minipage}
    \vspace{-0.2cm}
    \caption{DGP 4: Left: Five process realizations. Right: The densities of the estimated breakpoints for each specification.}
    \label{fig:Sim_4}
\end{figure}

\begin{figure}[H]
    \centering
    \begin{minipage}{0.38\textwidth}
        \centering
        \includegraphics[width=\textwidth]{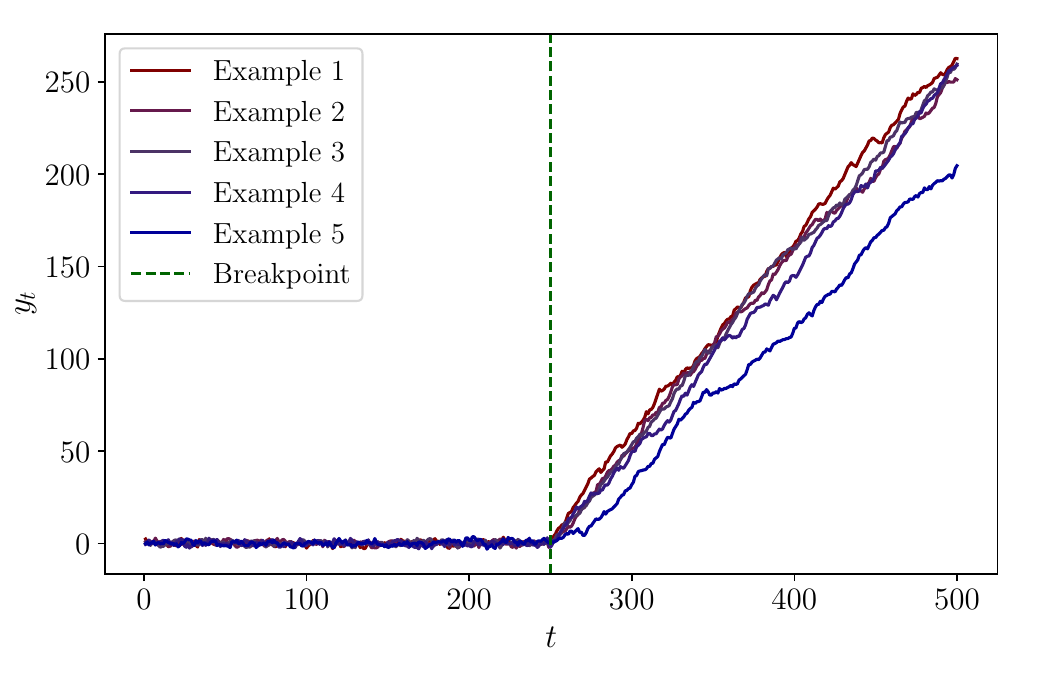}
    \end{minipage}%
    \hspace{0.5cm} 
    \begin{minipage}{0.38\textwidth}
        \centering
        \includegraphics[width=\textwidth]{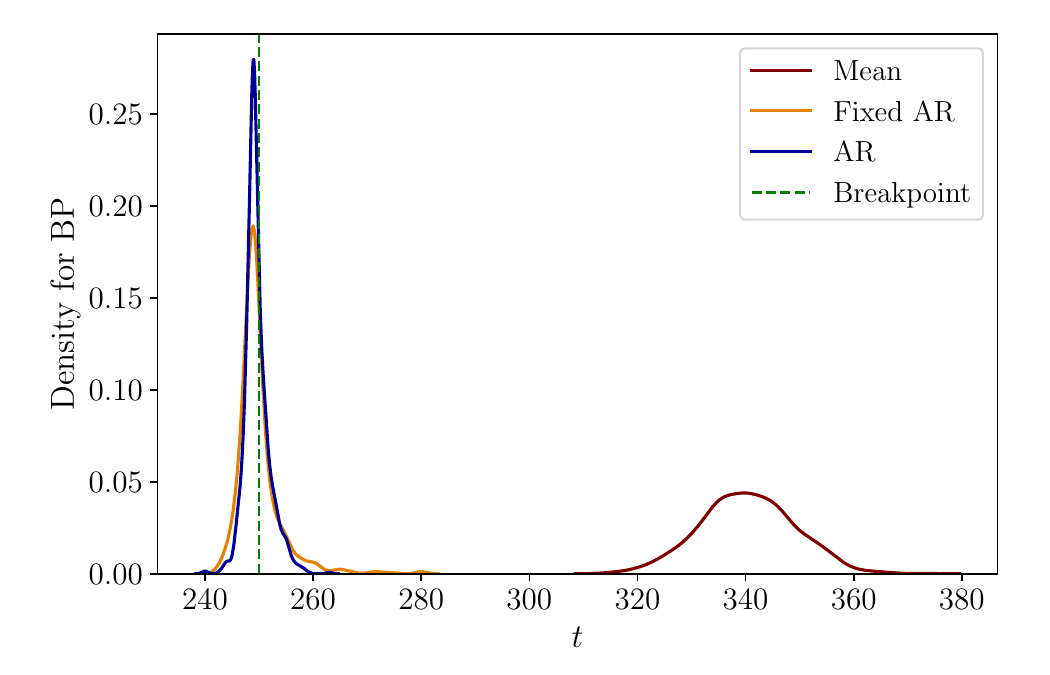}
    \end{minipage}
    \vspace{-0.2cm}
    \caption{DGP 5: Left: Five process realizations. Right: The densities of the estimated breakpoints for each specification.}
    \label{fig:Sim_5}
\end{figure}

\begin{figure}[H]
    \centering
    \begin{minipage}{0.38\textwidth}
        \centering
        \includegraphics[width=\textwidth]{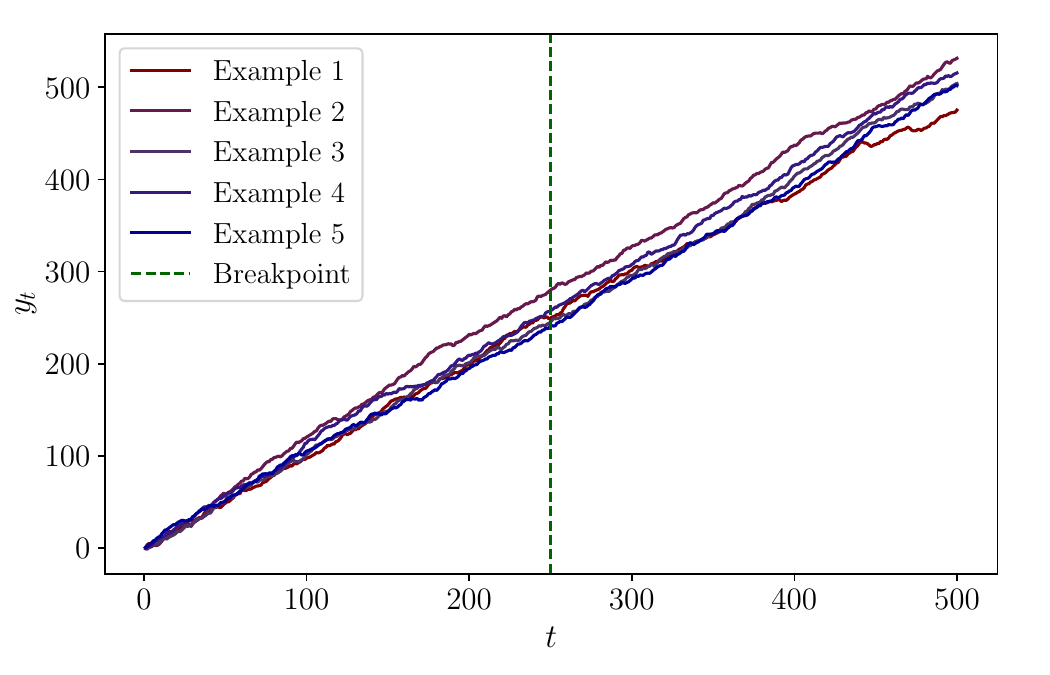}
    \end{minipage}%
    \hspace{0.5cm} 
    \begin{minipage}{0.38\textwidth}
        \centering
        \includegraphics[width=\textwidth]{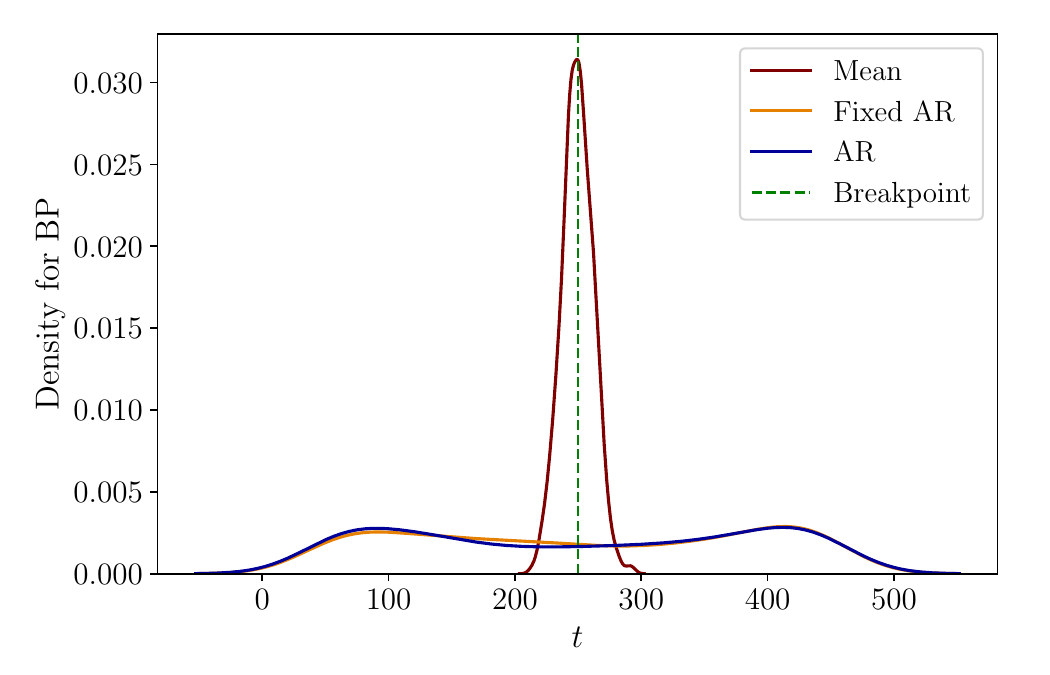}
    \end{minipage}
    \vspace{-0.2cm}
    \caption{DGP 6: Left: Five process realizations. Right: The densities of the estimated breakpoints for each specification.}
    \label{fig:Sim_6}
\end{figure}

\begin{figure}[H]
    \centering
    \begin{minipage}{0.38\textwidth}
        \centering
        \includegraphics[width=\textwidth]{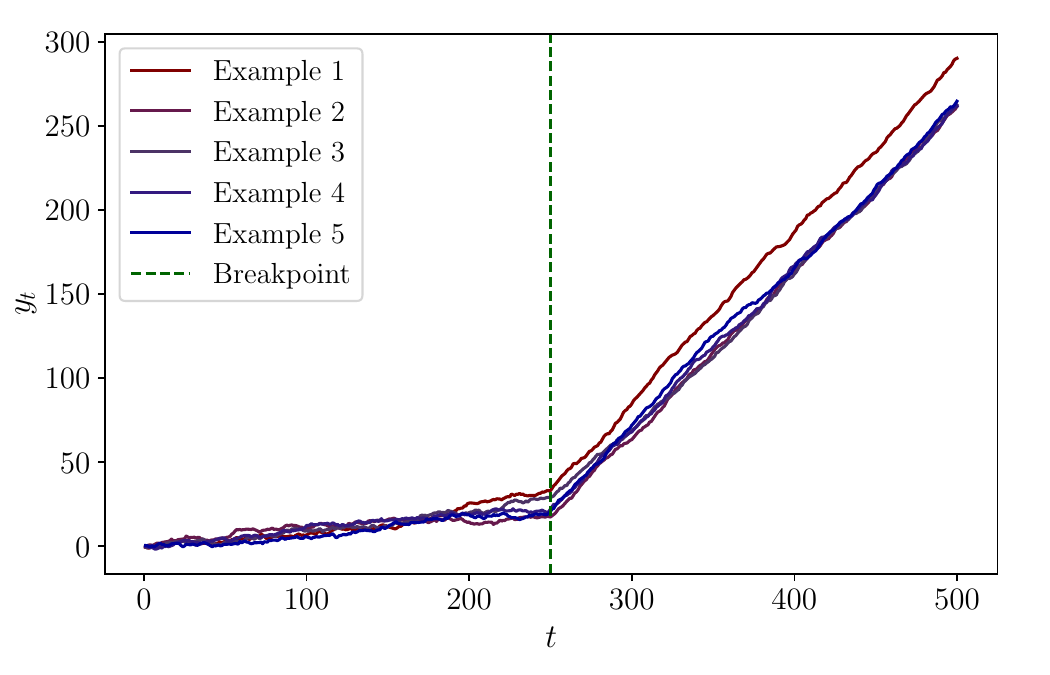}
    \end{minipage}%
    \hspace{0.5cm} 
    \begin{minipage}{0.38\textwidth}
        \centering
        \includegraphics[width=\textwidth]{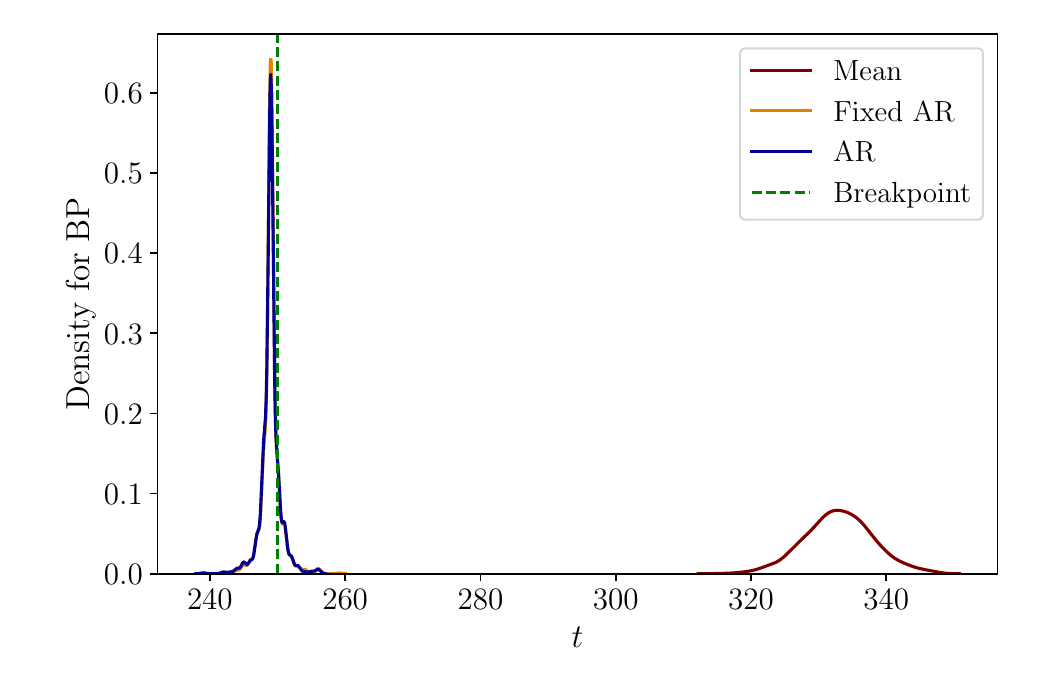}
    \end{minipage}
    \vspace{-0.2cm}
    \caption{DGP 7: Left: Five process realizations. Right: The densities of the estimated breakpoints for each specification.}
    \label{fig:Sim_7}
\end{figure}

\begin{figure}[H]
    \centering
    \begin{minipage}{0.38\textwidth}
        \centering
        \includegraphics[width=\textwidth]{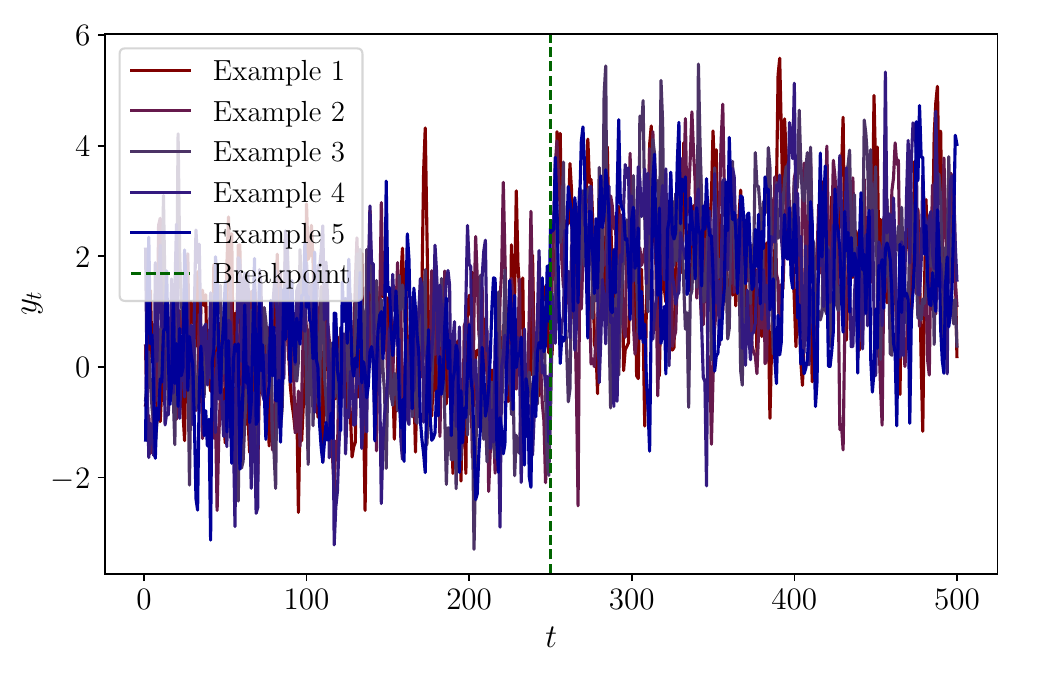}
    \end{minipage}%
    \hspace{0.5cm} 
    \begin{minipage}{0.38\textwidth}
        \centering
        \includegraphics[width=\textwidth]{fA8.1.pdf}
    \end{minipage}
    \vspace{-0.2cm}
    \caption{DGP 8: Left: Five process realizations. Right: The densities of the estimated breakpoints for each specification.}
    \label{fig:Sim_8}
\end{figure}

The DGPs range from random walk models with a break in the drift term to models with breaks in both the intercept and the AR coefficient. For comparison, we include a random walk without breakpoints as the sixth model. For each of the DGPs, we are interested in the performance of the methodology by \cite{BaiPerrson1998, BaiPerron2003} in estimating the breakpoint and confidence intervals. The model specifications from Section \ref{specification} are estimated on the data generated by the DGPs, and we use the implementation outlined in Section \ref{Implementation}. We use the R-package \textit{mbreaks} by \cite{R-mbreaks}, and we impose a single breakpoint in the estimation.
The left and right panels of Figs. \ref{fig:Sim_1} through \ref{fig:Sim_8} display realizations of the DGP and density plots of the estimated breakpoints for each of the models, respectively.
The results are summarized in Table \ref{tab:CI}, which provides the mean of the estimated breakpoints, and medians of the lower and upper boundaries of the estimated 95\% CIs are tabulated along with their coverage rates for each model and DGP.

\begin{table}[H]
\fontsize{9}{10}\selectfont
\centering 
\setlength{\tabcolsep}{3pt}
\begin{tabular}{ccccccccccccc}
\hline
DGP & \multicolumn{4}{c}{\uline{\quad\quad\quad\quad\quad\quad Mean \quad\quad\quad\quad\;\quad}} & \multicolumn{4}{c}{\uline{\quad\quad\quad\quad\quad Fixed AR \quad\quad\quad\quad\quad}} & \multicolumn{4}{c}{\uline{\quad\quad\quad\quad\quad\quad AR \quad\quad\quad\quad\quad\quad\;}} \\
      &  BP est. & Lower &  Upper &   Coverage &   BP est. & Lower &  Upper &   Coverage &  BP est. & Lower &  Upper &   Coverage  \\     
\hline
1 & 301 &  174 & 655 & 57.1\% & 251  & 216 & 336 & 43.4\% & 290 & 240 & 316 & 22.7\% \\
2 & 333 &  -386 & 332 & 95.4\% & 249  & 237 & 262 & 93\% &249  & 236 & 256 & 77.2\% \\
3 & 263 &  253 & 284 & 41.4\% & 256 & 239 & 260 & 89.9\% & 251 & 241 & 260 & 85.9\% \\
4 & 340 &  -190 & 340 & 97.5\% & 249 & 239 & 260 & 95.8\% & 249 & 238 & 250 & 65.8\% \\
5 & 340 &  -114 & 340 & 97.1\% & 250 & 239 & 258 & 97\% & 250 & 241 & 250 & 72.9\% \\
6 & 249 &  -3325 & 3976 & $\times$ & 253 & 142 & 371 & $\times$ & 254 & 202 & 312 & $\times$ \\
7 & 333 &  -282 & 330 & 92\% &249  & 246 & 253 & 97.8\% & 249 & 246 & 253 & 96\% \\
8 & 249 &  237 & 264 & 95.1\% & 248  & 236 & 263 & 95.2\% & 248 & 236 & 263 & 94.5\% \\
\hline
\end{tabular}
\vspace{-0.2cm}
 \caption{ Mean of the estimated breakpoints and medians of the lower and upper boundary of the estimated confidence intervals, along with the coverage rates for each model specification and DGP. DGP 6 is simulated without a breakpoint, so the coverage rate is irrelevant and indicated by $\times$.}
    \label{tab:CI}
\end{table}

\vspace{-0.3cm}

In the first DGP, a random walk with a small drift term break, we observe that the mean of the estimated breakpoints is later than the true breakpoint in all model specifications. Additionally, the density plots exhibit asymmetry around the true breakpoint. This is expected due to the low magnitude of the break in the drift term, which creates a subtle change in the overall stochastic trend, making accurate breakpoint detection difficult. In the second DGP with a larger drift term break, the estimated breakpoints exhibit a narrower and more bell-shaped density. The mean estimated breakpoints for the Fixed AR and AR models slightly precede the true breakpoint. However, the Mean model performs poorly, with the mean of the estimated breakpoints far from the true breakpoint.  

In the third DGP, both the Fixed AR and AR models produce mean estimated breakpoints slightly later than the true breakpoint. The Mean model exhibits better performance in this DGP than in the second DGP. The fourth DGP has a break in the intercept and the AR-coefficient from 0.95 to 1, resulting in a state-wise non-stationary model. This change leads to breakpoint estimates very close to the true breakpoint, except in the Mean model. A similar outcome is observed in the fifth DGP, which features a larger increase in the AR-coefficient. In the sixth DGP, which is defined without any breakpoints, the Mean model estimates breakpoints near the midpoint of the sample period, while the other two specifications yield inconclusive results. In the seventh DGP, the AR and Fixed AR models produce estimates close to the true breakpoint. However, the Mean model continues to produce breakpoint estimates far from the true value. Examining the eighth DGP, the three models perform almost equally well.

Overall, the Fixed AR and AR models tend to perform well in non-stationary scenarios, estimating breakpoints close to the true breakpoints. The methodology, however, appears to struggle with accurately estimating the true breakpoint in cases of minor changes between states and large error term variance.
In contrast, the Mean model does not perform well in DGPs featuring gradual changes, aligning with theoretical expectations as detailed in \cite{BaiPerron2003}.

The coverage rate of a CI is the proportion of times the CI covers the true breakpoint, here at $t=250$. We find that the CIs of the Mean model are generally very wide and have varying coverage. In the Fixed AR and AR models, the CIs are typically narrower. The coverage rates are best in the DGPs with large differences between the states as seen in DGPs 4, 5, 7 and 8 using the Fixed AR model specification, which is in line with the findings of \cite{BaiPerron2003}. For the AR model, the coverage rates are only close to the desired 95\% in the seventh and eighth DGP, indicating that the CIs are inadequate in most of the DGPs considered.

\begin{table}[H]
\fontsize{9}{10}\selectfont
\centering
\setlength{\tabcolsep}{3pt} 
\begin{tabular}{cccccccccc}
\hline
DGP & \multicolumn{3}{c}{\uline{\quad\quad\quad\quad\quad\quad Mean \quad\quad\quad\quad\;\quad}} & \multicolumn{3}{c}{\uline{\quad\quad\quad\quad\quad Fixed AR \quad\quad\quad\quad\quad}} & \multicolumn{3}{c}{\uline{\quad\quad\quad\quad\quad\quad\quad AR \quad\quad\quad\quad\quad\quad}} \\
      &  BIC &  LWZ &   KT  &  BIC &  LWZ &   KT  &  BIC &  LWZ &   KT \\     
\hline
1 & 3.0 (0\%) & 3.0 (0\%) & 3.0 (0\%)  & 0.2 (15\%) & 0.0 (0\%) & 3.0 (0\%)    & 0.1 (6\%) & 0.0 (0\%) & 0.0 (3\%)  \\
2 & 3.0 (0\%) & 3.0 (0\%) & 3.0 (0\%)  & 1.0 (97\%) & 0.8 (82\%) & 3.0 (0\%)   & 1.0 (94\%) & 0.5 (46\%) & 1.0 (93\%)  \\
3 & 2.9 (0\%) & 2.7 (4\%) & 3.0 (0\%)  & 1.0 (94\%) & 0.2 (16\%) & 2.9 (0\%)   & 0.9 (85\%) & 0.0 (0\%) & 0.7 (70\%) \\
4 & 3.0 (0\%) & 3.0 (0\%) & 3.0 (0\%)  & 1.0 (98\%) & 1.0 (98\%) & 2.8 (0\%)   & 1.0 (99\%) & 0.9 (92\%) & 1.0 (99\%) \\
5 & 3.0 (0\%) & 3.0 (0\%) & 3.0 (0\%)  & 1.0 (99\%) & 1.0 (97\%) & 2.7 (0\%)   & 1.0 (99\%) & 1.0 (100\%) & 1.0 (99\%)  \\
6 & 3.0 (0\%) & 3.0 (0\%) & 3.0 (0\%)  & 0.0 (98\%) & 0.0 (100\%) & 3.0 (0\%)  & 0.0 (100\%) & 0.0 (100\%) & 0.0 (100\%) \\
7 & 3.0 (0\%) & 3.0 (0\%) & 3.0 (0\%)  & 1.0 (99\%) & 1.0 (100\%) & 3.0 (0\%)  & 1.0 (98\%) & 1.0 (100\%) & 1.0 (98\%)  \\
8 & 1.5 (63\%) & 1.0 (98\%) & 1.3 (72\%)& 1.0 (99\%) & 1.0 (100\%) & 1.3 (73\%)& 1.0 (100\%) & 1.0 (98\%) & 1.0 (100\%) \\

\hline
\end{tabular}
\vspace{-0.2cm}
\caption{Means of the estimated number of breakpoints for each model specification across different DGPs, rounded to one decimal. Percentages indicate the proportion of estimates equal to the true number of breakpoints.}
    \label{tab:Mean_est_BP}
\end{table}
\vspace{-0.3cm}

Table \ref{tab:Mean_est_BP} shows the mean number of breakpoints estimated for each DGP and method, along with the proportion of correctly estimated breakpoints.
The difficulty in accurately estimating gradual changes using the Mean model is also evident when estimating the number of breakpoints. This model specification leads to overestimating the number of breakpoints in all DGPs considered except DGP 8, where it performs well. The BIC criterion in the Fixed AR specification performs very well, with an estimated number of breakpoints equal to the true number in most simulations in DGP 2-8. The LWZ criterion performs almost equally well except in the third DGP, while the KT criterion vastly overestimates the number of breakpoints in DGP 1-7. In the AR model, the information criteria all perform well in DGPs 2-8 except for the third DGP where the LWZ criterion underestimates the number of breakpoints.

\vspace{-0.5cm}

\vspace{-0.5cm}

\vspace{-0.5cm}

\vspace{-0.5cm}

\vspace{-0.5cm}

\vspace{-0.5cm}

\vspace{-0.5cm}

\subsection{Serially correlated error term}
\label{simulation_study_serial}
    
A possible extension of the simulation study outlined in Eq. \eqref{sim_model} is allowing the error term to exhibit serial correlation. We use the same DGPs as before, but generate $\{\varepsilon_t\}_{t=1}^T$ as follows,
\begin{align}
    \varepsilon_t &= \psi \varepsilon_{t-1} + \theta \eta_{t-1}+\eta_t, \quad \eta_t \overset{\textit{i.i.d.}}{\sim} \mathcal{N}(0, \sigma_{\eta}^2) \quad \forall t.
\end{align}

We conduct 1000 simulations for each, with a sample size of 500. Here, we consider DGPs 2, 3, 4, 5, 7, and 8 as outlined in Table \ref{table:sim_specification} and refer to these DGPs in the serially correlated cases as models $2_s$, $3_s$, $4_s$, $5_s$, $7_s$, and $8_s$. We set $\psi=\theta=0.5$ and the standard deviation $\sigma_{\eta}$, such that the standard deviation of $\varepsilon_t$ corresponds to the $\sigma$ in Table \ref{table:sim_specification}. This is accomplished as follows,
\begin{align}
    \operatorname{Var}\left(  \varepsilon_t\right)&= \operatorname{Var}\left( \psi \varepsilon_{t-1} + \theta \eta_{t-1}+\eta_t \right) \nonumber\\
    &= \psi^2 \operatorname{Var} \left(\varepsilon_{t-1}\right) + \theta^2  \operatorname{Var} \left(\eta_{t-1}\right) + 2\psi \theta \operatorname{Cov}\left(\varepsilon_{t-1}, \eta_{t-1}\right) +  \operatorname{Var} \left(\eta_{t}\right).  \nonumber\\
    &=  \psi^2 \operatorname{Var} \left(\varepsilon_{t-1}\right) + \theta^2  \sigma_{\eta}^2 + 2\psi \theta \sigma_{\eta}^2 +  \sigma_{\eta}^2, \nonumber
    \intertext{since $\varepsilon_{t-1}$ and $\eta_{t-1}$ have zero means and $\mathbb{E}\left[\varepsilon_t \eta_t\right]=\phi \mathbb{E}\left[\varepsilon_{t-1}\eta_t \right]+ \theta \mathbb{E}\left[\eta_t \eta_{t-1}\right]+ \mathbb{E}\left[\eta_t^2\right]=\sigma_{\eta}^2$. Given stationarity of the process, which implies $\sigma^2=  \operatorname{Var}\left(  \varepsilon_t\right)$ for all $t$, we derive,}
    \sigma_{\eta}^2 &= \sigma^2 \frac{1-\psi^2}{1+ \theta^2 + 2 \psi \theta}. \nonumber
\end{align}
This adjustment ensures the comparability of the results between the two error term types.

\begin{figure}[H]
    \fontsize{10}{11}\selectfont
    \centering
    \begin{minipage}{0.38\textwidth}
        \centering
        \includegraphics[width=\textwidth]{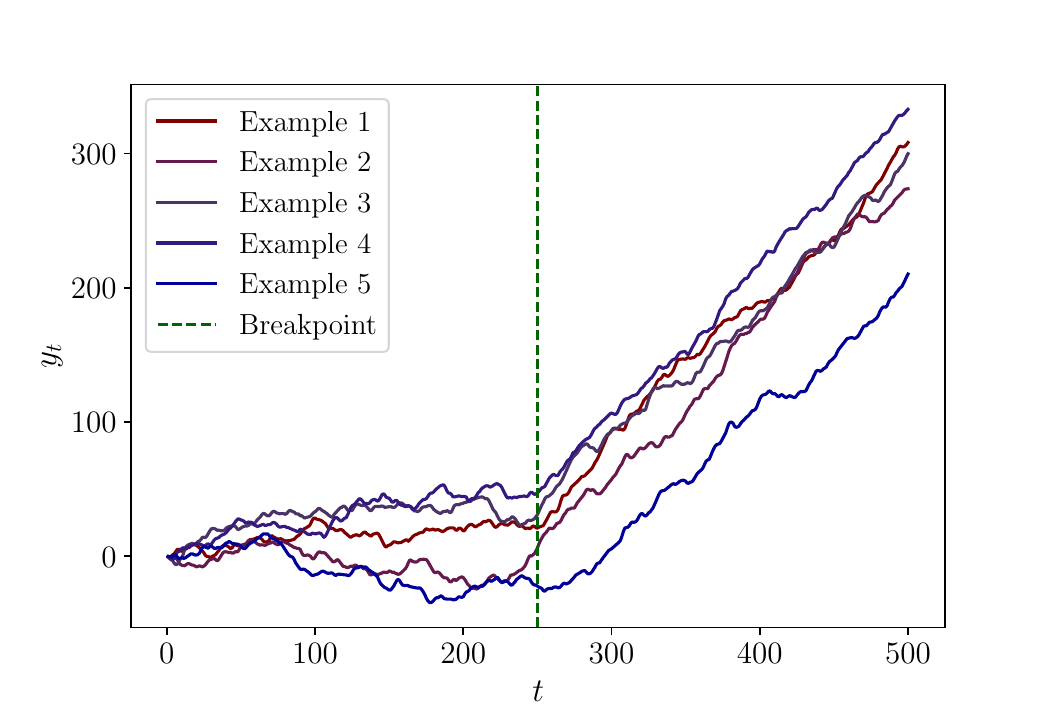}
    \end{minipage}%
    \hspace{0.5cm}
    \begin{minipage}{0.38\textwidth}
        \centering
        \includegraphics[width=\textwidth]{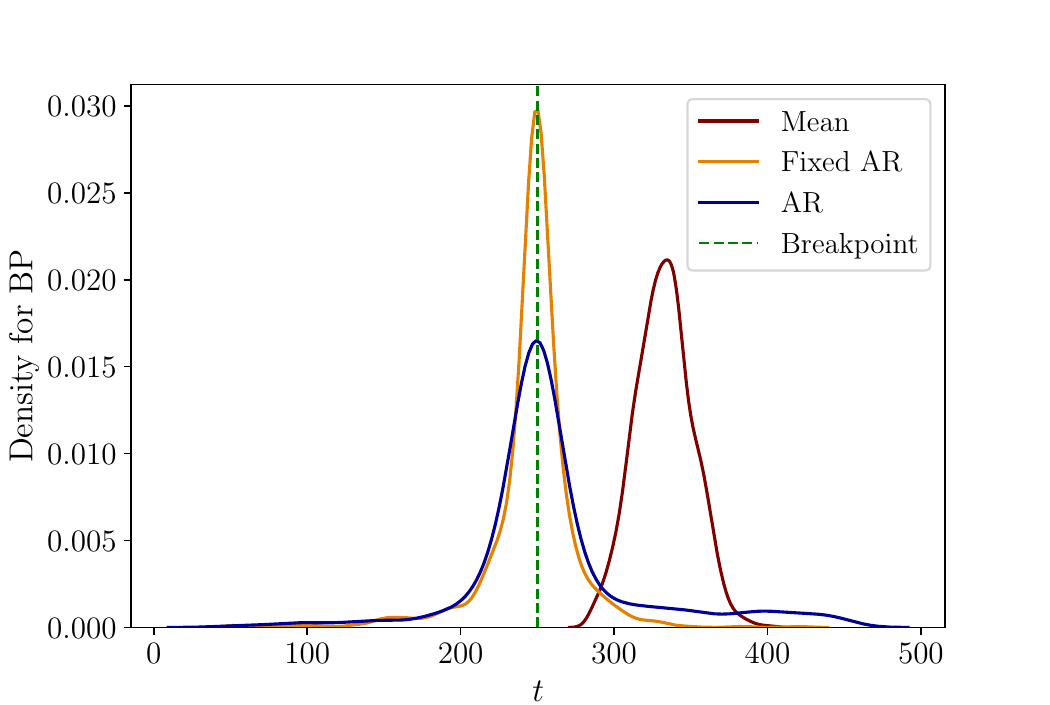}
    \end{minipage}
       \vspace{-0.2cm}
    \caption{DGP $2_s$: Left: Five process realizations. Right: The densities of the estimated breakpoints for each specification.}
    \label{fig:ARMASim_2}
\end{figure}

\begin{figure}[H]
    \centering
    \begin{minipage}{0.38\textwidth}
        \centering
        \includegraphics[width=\textwidth]{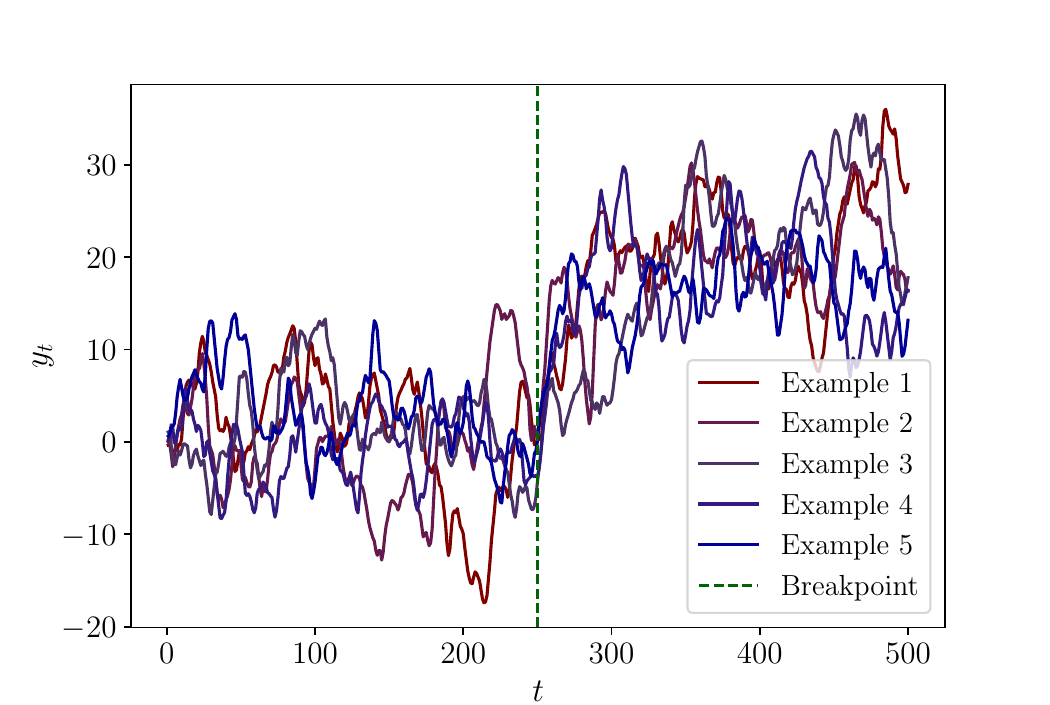}
    \end{minipage}%
    \hspace{0.5cm}
    \begin{minipage}{0.38\textwidth}
        \centering
        \includegraphics[width=\textwidth]{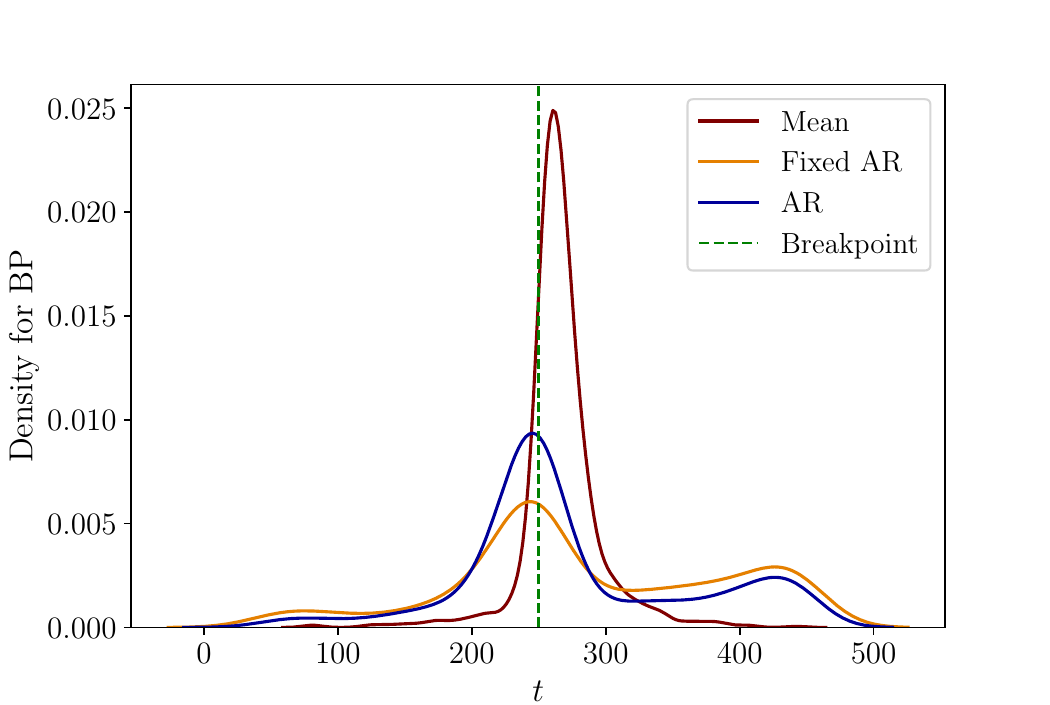}
    \end{minipage}
       \vspace{-0.2cm}
    \caption{DGP $3_s$: Left: Five process realizations. Right: The densities of the estimated breakpoints for each specification.}
    \label{fig:ARMASim_3}
\end{figure}

\begin{figure}[H]
    \centering
    \begin{minipage}{0.38\textwidth}
        \centering
        \includegraphics[width=\textwidth]{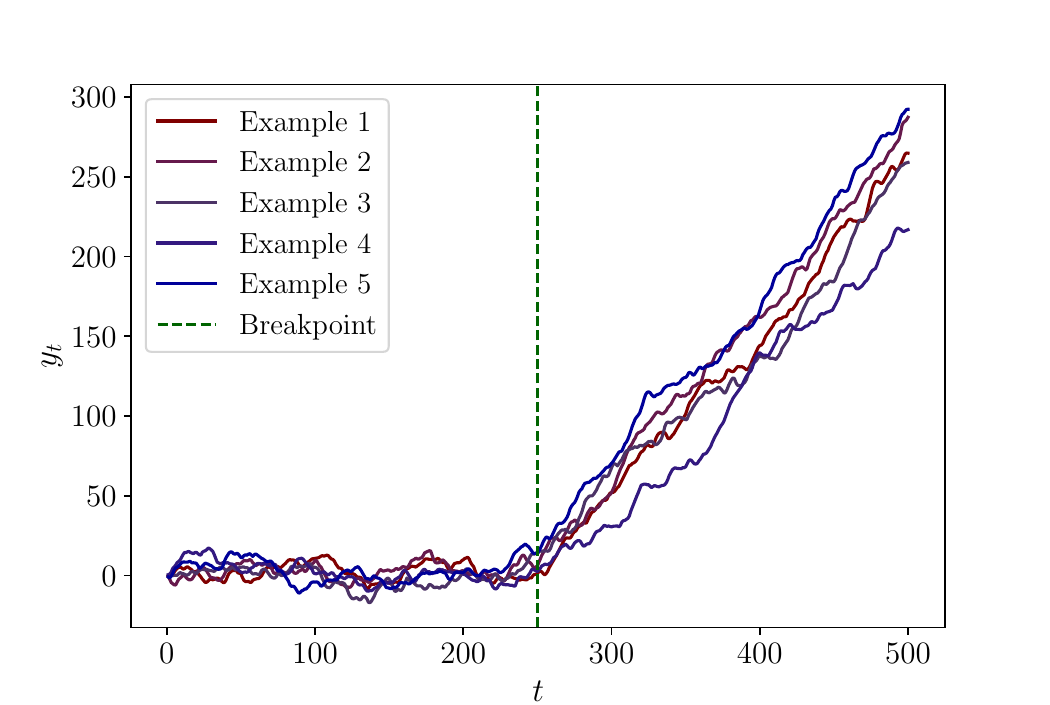}
    \end{minipage}%
    \hspace{0.5cm}
    \begin{minipage}{0.38\textwidth}
        \centering
        \includegraphics[width=\textwidth]{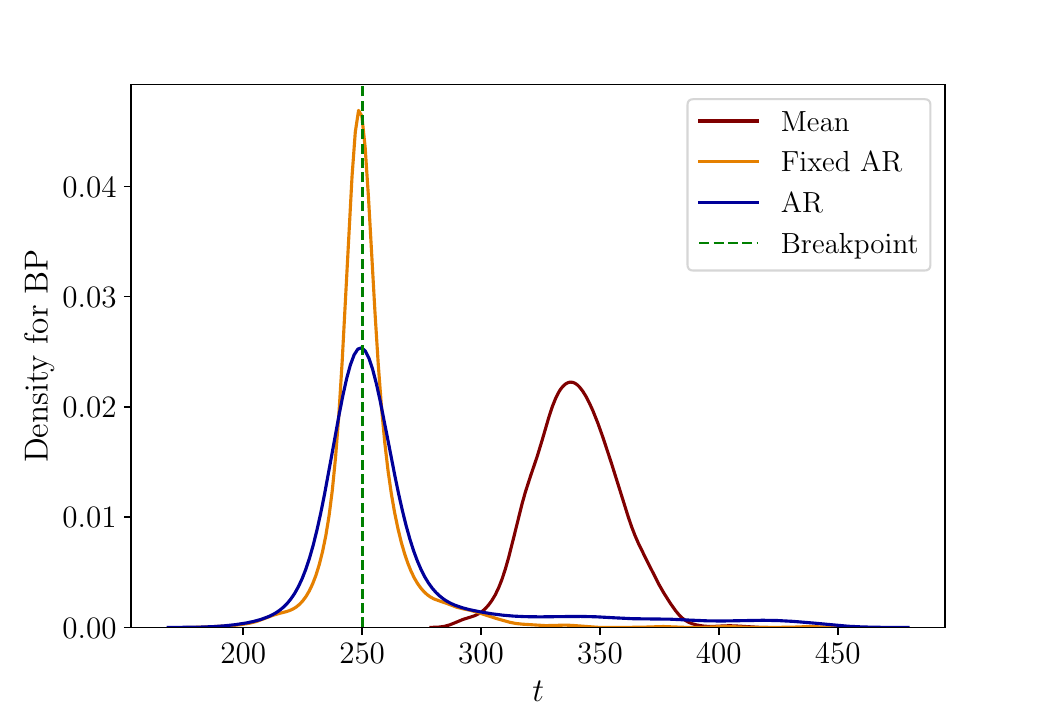}
    \end{minipage}
       \vspace{-0.2cm}
    \caption{DGP $4_s$: Left: Five process realizations. Right: The densities of the estimated breakpoints for each specification.}
    \label{fig:ARMASim_4}
\end{figure}

\begin{figure}[H]
    \centering
    \begin{minipage}{0.38\textwidth}
        \centering
        \includegraphics[width=\textwidth]{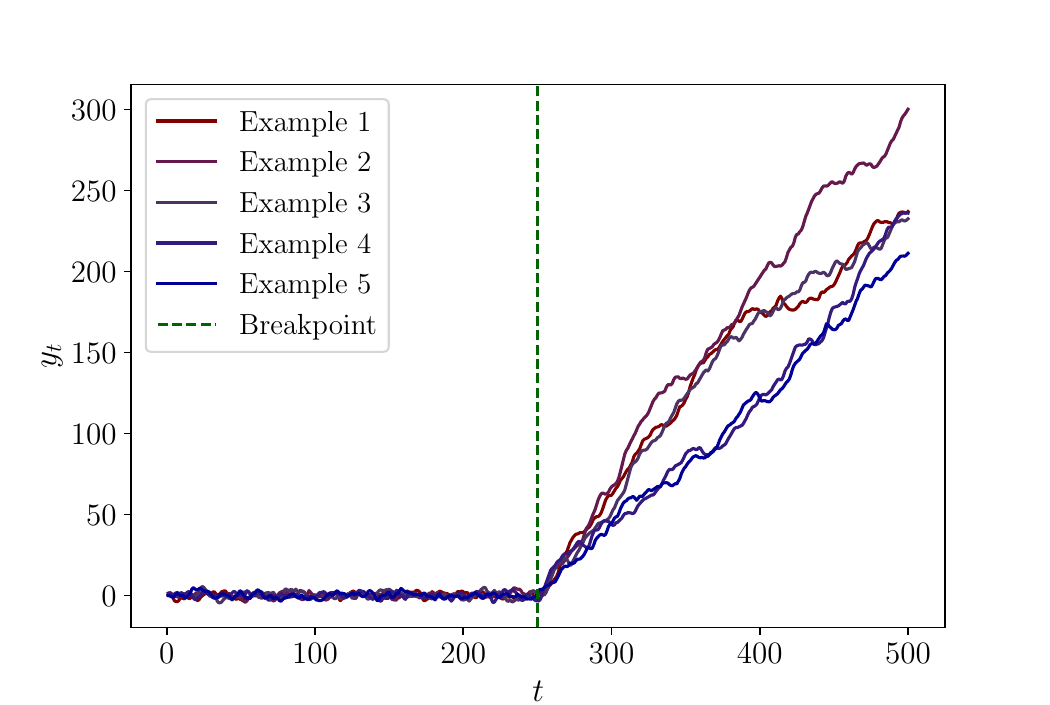}
    \end{minipage}%
    \hspace{0.5cm}
    \begin{minipage}{0.38\textwidth}
        \centering
        \includegraphics[width=\textwidth]{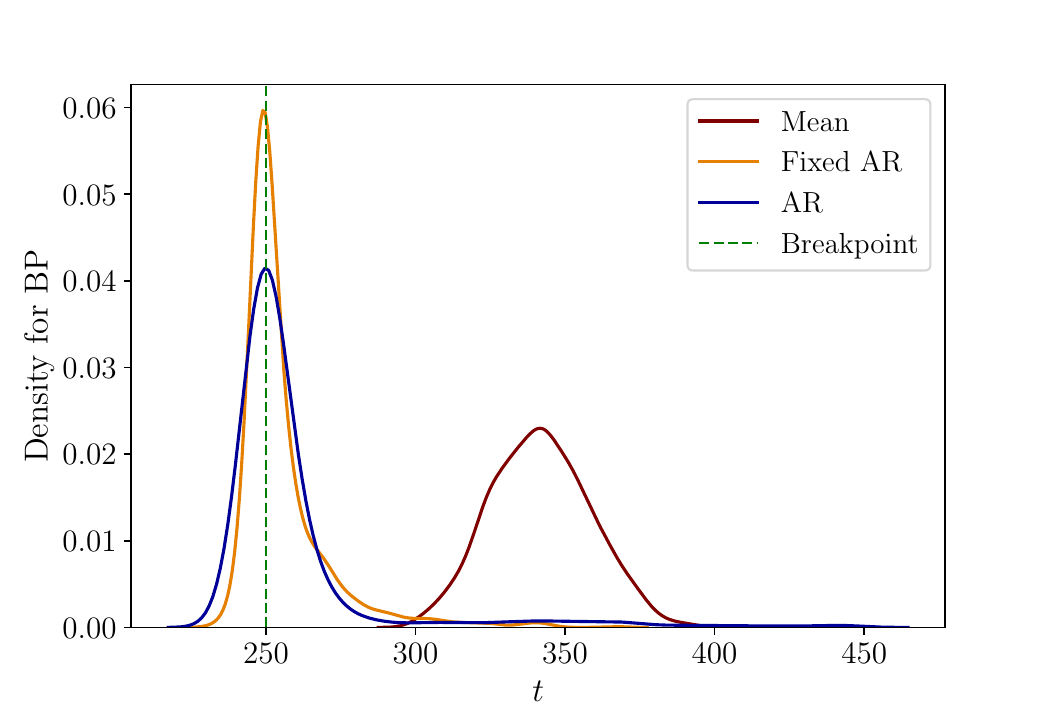}
    \end{minipage}
       \vspace{-0.2cm}
    \caption{DGP $5_s$: Left: Five process realizations. Right: The densities of the estimated breakpoints for each specification.}
    \label{fig:ARMASim_5}
\end{figure}

\begin{figure}[H]
    \centering
    \begin{minipage}{0.38\textwidth}
        \centering
        \includegraphics[width=\textwidth]{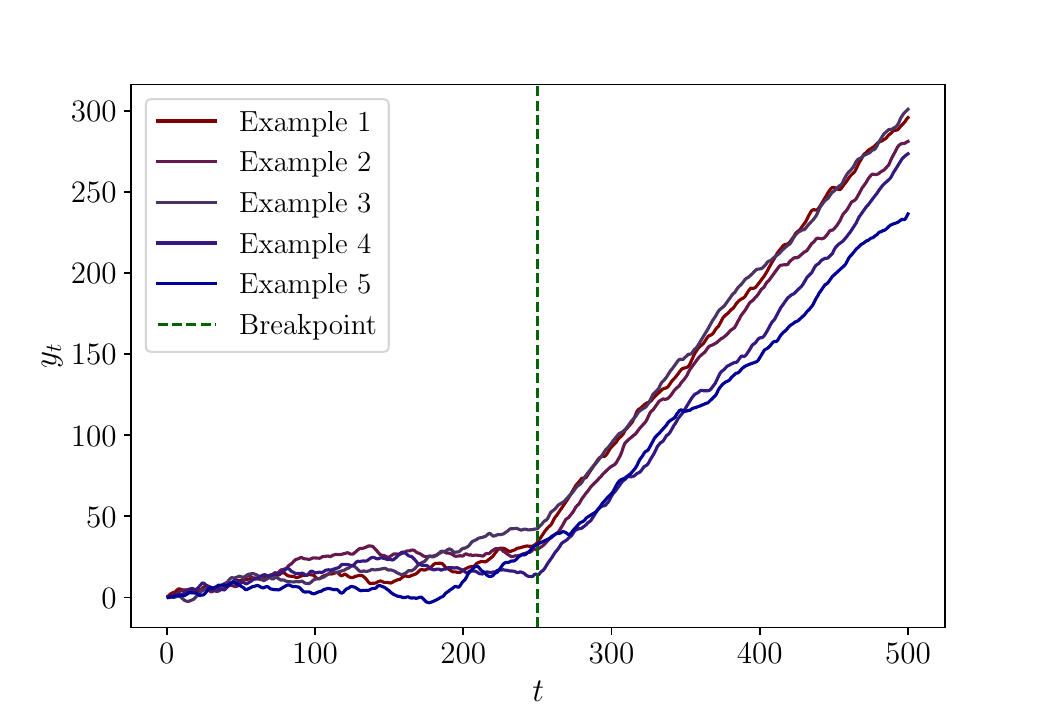}
    \end{minipage}%
    \hspace{0.5cm}
    \begin{minipage}{0.38\textwidth}
        \centering
        \includegraphics[width=\textwidth]{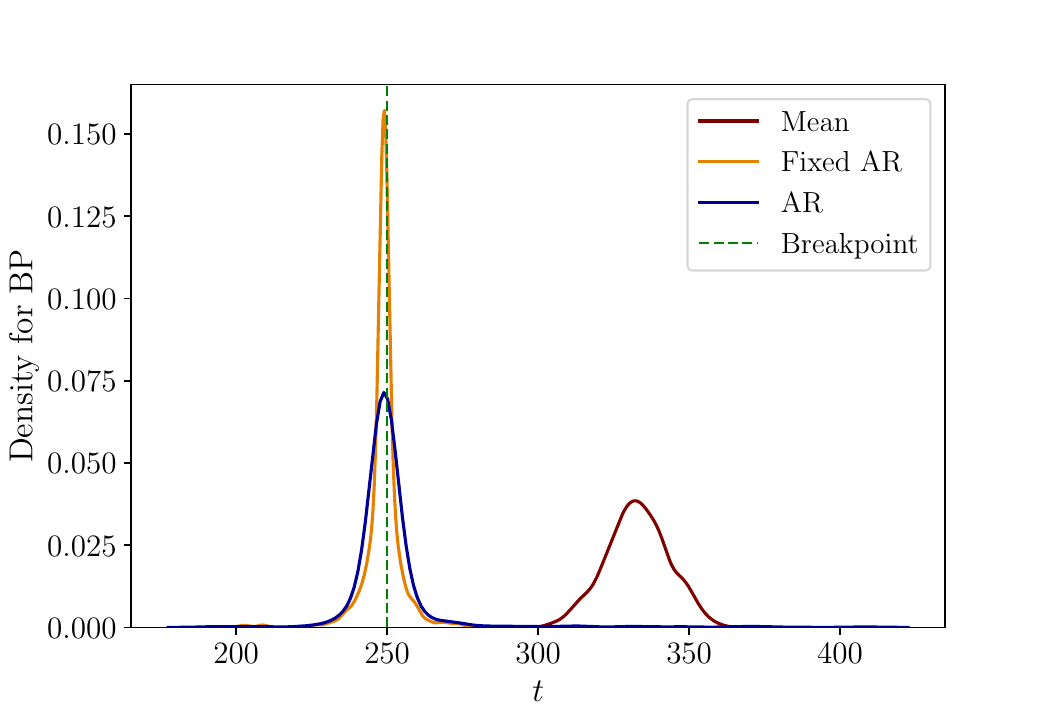}
    \end{minipage}
       \vspace{-0.2cm}
    \caption{DGP $7_s$: Left: Five process realizations. Right: The densities of the estimated breakpoints for each specification.}
    \label{fig:ARMASim_7}
\end{figure}

\begin{figure}[H]
    \centering
    \begin{minipage}{0.38\textwidth}
        \centering
        \includegraphics[width=\textwidth]{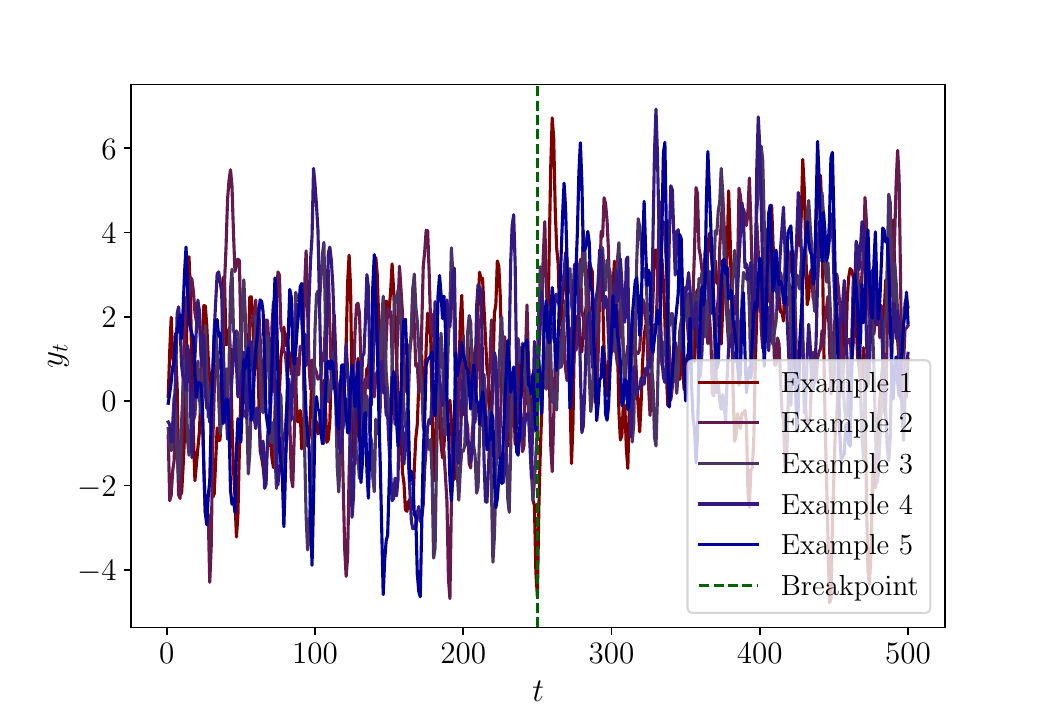}
    \end{minipage}%
    \hspace{0.5cm}
    \begin{minipage}{0.38\textwidth}
        \centering
        \includegraphics[width=\textwidth]{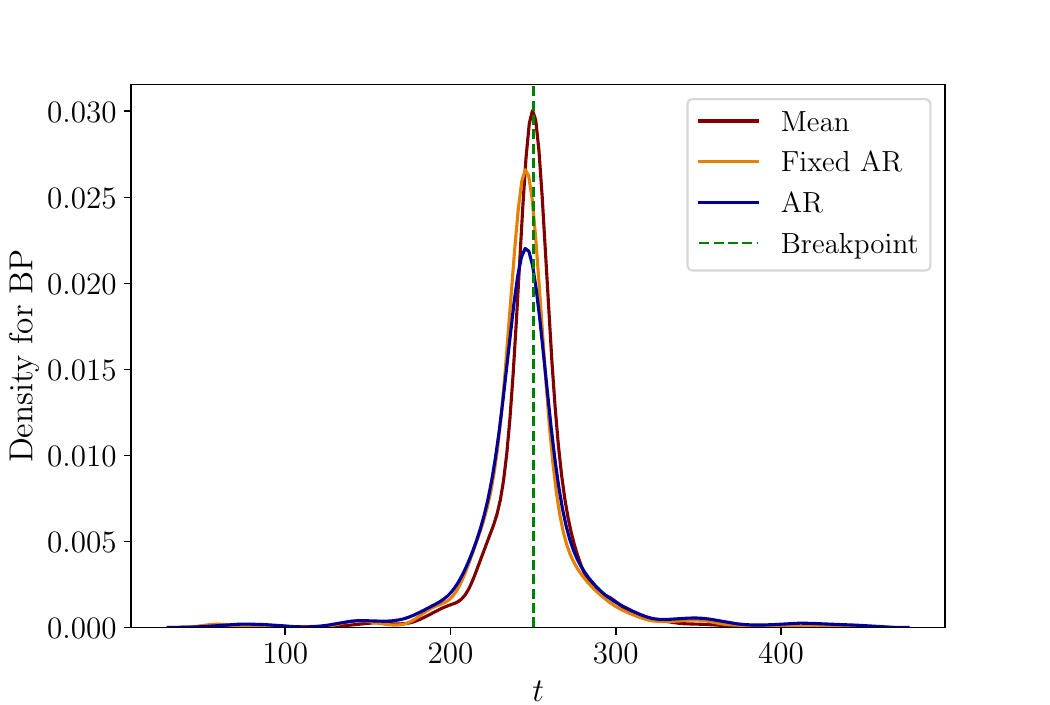}
    \end{minipage}

    \caption{DGP $8_s$: Left: Five process realizations. Right: The densities of the estimated breakpoints for each specification.}
    \label{fig:ARMASim_8}
\end{figure}

In Figs. \ref{fig:ARMASim_2} through \ref{fig:ARMASim_8}, we plot examples of realizations and frequency plots of the estimated breakpoints using each of the models while imposing a single breakpoint in the estimation. 
The results are summarized in Table \ref{tab:CI_ARMA}, which provides means of the estimated breakpoints and medians of the lower and upper boundary of the estimated confidence intervals, along with the coverage rates for each model specification and DGP. Generally speaking, the mean of the estimated breakpoints are further from the true breakpoint and the CIs become wider compared to the results from the corresponding DGPs without serial correlation. 
It is evident that serial correlation in the error term makes it more difficult to estimate the dating of breaks. We find that the Fixed AR and AR models perform well for DGP $7_s$, which has a large difference between the states and low variance. This is in line with the theoretical framework by \cite{BaiPerron2003}, who note that the estimated break dates are consistent even in the presence of serial correlation. The Fixed AR model performs well in DGPs $2_s$, $4_s$ and $5_s$ where the mean of the estimated breakpoints is close to the true breakpoint, and confidence intervals are reasonably wide with acceptable coverage rates. The results of the AR model are less conclusive. 

For the Mean and Fixed AR models, the coverage rates are generally close to the desired 95\% and even higher in some DGPs. However, the CIs are also extremely wide, reaching outside the sample window in many DGPs. The CIs seem reasonable in the Fixed AR model for DGPs $2_s$, $4_s$, $5_s$, and $7_s$, where the coverage rates are close to 95\% and the medians of the lower and upper bounds of the CIs are not too extreme. The CIs for the AR model are generally wider than in the version without serial correlation in the error term. In the AR model, the coverage rates are lower than the desired 95\%, but it seems that DGPs with large breaks have higher coverage rates. 
The relatively poor performance is in line with the theoretical framework by \cite{BaiPerron2003}. The authors note that the construction of the CIs rely on having no serial correlation in the error term if a lagged dependent variable is included as a regressor that has coefficients that are subject to breakpoints.
  
\begin{table}[H]
\fontsize{9}{10}\selectfont
\centering 
\setlength{\tabcolsep}{3pt}

\begin{tabular}{ccccccccccccc}
\hline
DGP & \multicolumn{4}{c}{\uline{\quad\quad\quad\quad\quad\quad Mean \quad\quad\quad\quad\;\quad}} & \multicolumn{4}{c}{\uline{\quad\quad\quad\quad\quad Fixed AR \quad\quad\quad\quad\quad}} & \multicolumn{4}{c}{\uline{\quad\quad\quad\quad\quad\quad AR \quad\quad\quad\quad\quad\quad\;}} \\
      & BP est. & Lower &  Upper &   Coverage & BP est. &  Lower &  Upper &   Coverage & BP est. & Lower &  Upper &   Coverage  \\     
\hline
$2_s$ & 332  & -1400 & 335 & 95.9\%  & 247 & 188 & 312 & 95.7\% & 261 & 190 & 299 & 79.9\% \\
$3_s$ & 266 & 60 & 787 & 90.6\%      & 285 & -112 & 656 & 97.2\% & 276 & 156 & 421 & 77.1\% \\
$4_s$ & 340 & -776 & 339 & 94.9\%    & 252 & 197 & 301 & 96.9\% & 264  & 195 & 277 & 84.9\% \\
$5_s$ & 342 & -329 & 340 & 96.2\%    & 256 & 196 & 266 & 96.4\% & 259 & 192 & 250 & 70.8\% \\
$7_s$ & 333 & -1708 & 329 & 92.3\%   & 249 & 230 & 270 & 97.6\% & 251 & 230 & 267 & 92.8\% \\
$8_s$ & 250 & 122 & 370 & 98.3\%     & 245 & -5 & 492 & 99.8\% & 247 & 23 & 490 & 97.4\% \\
\hline
\end{tabular}
\vspace{-0.2cm}
 \caption{ Mean of the estimated breakpoints and medians of the lower and upper boundary of the estimated confidence intervals, along with the coverage rates for each model specification and DGP.}
    \label{tab:CI_ARMA}
\end{table}

Table \ref{tab:Mean_est_BP_ARMA} shows the mean number of breakpoints estimated for each DGP and method, along with the proportion of correctly estimated number. In the Mean model, all information criteria overestimate the number of breakpoints. An important exception is the eighth DGP, where the performance is better, as in the case without serial correlation. 
In the Fixed AR and AR model specifications, the LWZ criterion generally performs well, while both the BIC and the KT criteria generally overestimate. 
However, the LWZ criterion leads to underestimating the number of breakpoints in DGPs $3_s$ and $8_s$. These two DGPs are characterized by fixed AR-coefficients that are lower than one. This implies that these two processes do not exhibit an autoregressive unit root. Hence, it seems that the LWZ criterion performs well in cases of state-wise non-stationarity or switching between stationary and non-stationary states.

Compared to the findings in the DGPs without serial correlation, it is clear that the proportion of correct estimates are lower for most DGPs and model specifications.
Overall, the best performing criterion seems to be the LWZ criterion in the Fixed AR and AR models, while the Mean model typically leads to overestimating the number of breakpoints.

\begin{table}[H]
\fontsize{9}{10}\selectfont
\centering
\setlength{\tabcolsep}{3pt} 
\begin{tabular}{cccccccccc}
\hline
DGP & \multicolumn{3}{c}{\uline{\quad\quad\quad\quad\quad\quad Mean \quad\quad\quad\quad\;\quad}} & \multicolumn{3}{c}{\uline{\quad\quad\quad\quad\quad Fixed AR \quad\quad\quad\quad\quad}} & \multicolumn{3}{c}{\uline{\quad\quad\quad\quad\quad\quad\quad AR \quad\quad\quad\quad\quad\quad}} \\
      &  BIC &  LWZ &   KT &   BIC &  LWZ &   KT &  BIC &  LWZ &   KT \\     
\hline
$2_s$ & 3.0 (0\%) & 3.0 (0\%) & 3.0 (0\%) & 1.9 (32\%) & 0.9 (70\%) & 2.9 (0\%)  & 1.8 (37\%) & 0.7 (61\%) & 1.9 (33\%)  \\
$3_s$ & 3.0 (0\%) & 2.8 (2\%) & 3.0 (0\%) & 0.7 (33\%) & 0.0 (0\%) & 2.7 (3\%) & 0.3 (19\%) & 0.0 (0\%) & 0.4 (17\%)  \\
$4_s$ & 3.0 (0\%) & 3.0 (0\%) & 3.0 (0\%) & 1.7 (45\%) & 1.0 (85\%) & 2.8 (1\%) & 1.6 (51\%) & 0.8 (79\%) & 1.6 (47\%) \\
$5_s$ & 3.0 (0\%) & 3.0 (0\%) & 3.0 (0\%) & 1.8 (5\%) & 1.1 (85\%) & 2.8 (0\%)  & 1.7 (40\%) & 1.0 (92\%) & 1.6 (49\%)  \\
$7_s$ & 3.0 (0\%) & 3.0 (0\%) & 3.0 (0\%) &  1.9 (34\%) & 1.1 (89\%) & 3.0 (0\%)  & 1.9 (34\%) & 1.0 (96\%) & 1.9 (32\%)  \\
$8_s$ & 2.2 (21\%)& 1.2 (78\%)& 2.2 (23\%)& 0.4 (35\%) & 0.0 (0\%) & 1.9 (36\%)  & 0.0 (4\%) & 0.0 (0\%) & 0.0 (3\%)   \\
\hline
\end{tabular}
   \vspace{-0.2cm}
 \caption{ Means of the estimated number of breakpoints for each model specification across different DGPs, rounded to one decimal. Percentages indicate the proportion of estimates equal to the true number of breakpoints.}
    \label{tab:Mean_est_BP_ARMA}
\end{table}

\vspace{-0.5cm}

\vspace{-0.5cm}

\vspace{-0.5cm}

\vspace{-0.5cm}

\vspace{-0.5cm}

       \vspace{0.6cm}

\section{Graphs}

\subsection{Reversed time}

\label{Reversed}

\begin{figure}[H]
    \centering
        \includegraphics[width=0.8\textwidth]{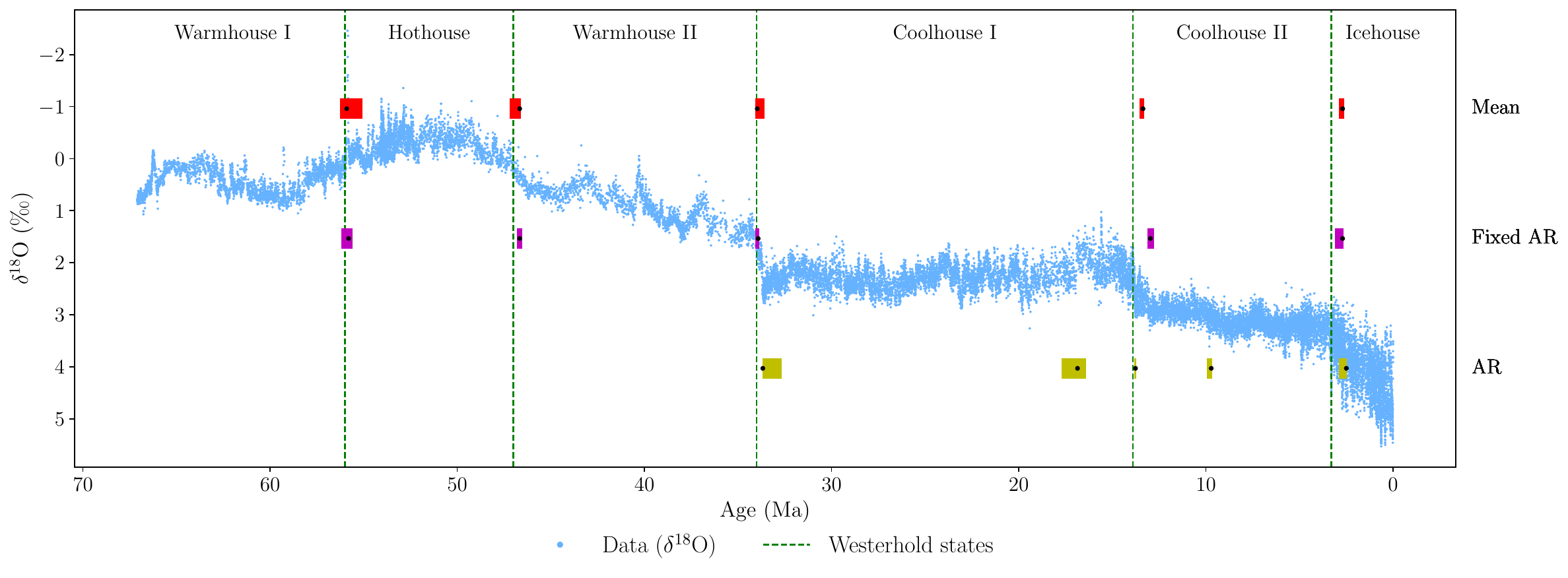}

    \caption{
A comparison of estimated breakpoints using the Mean, Fixed AR, and AR model specifications for five breakpoints on 25 kyr binned data where the time frame is reversed. The black dots represent estimated breakpoints, while colored shaded rectangles indicate 95\% confidence intervals. The results overlay the $\delta^{18}$O data from \cite{Westerhold2020} and their climate states. }
    \label{fig:reversed}
\end{figure}

\subsection{One to fifteen breakpoints: Mean model}
\label{1_15_Mean}

\begin{figure}[H]
    \centering
        \includegraphics[width=0.8\textwidth]{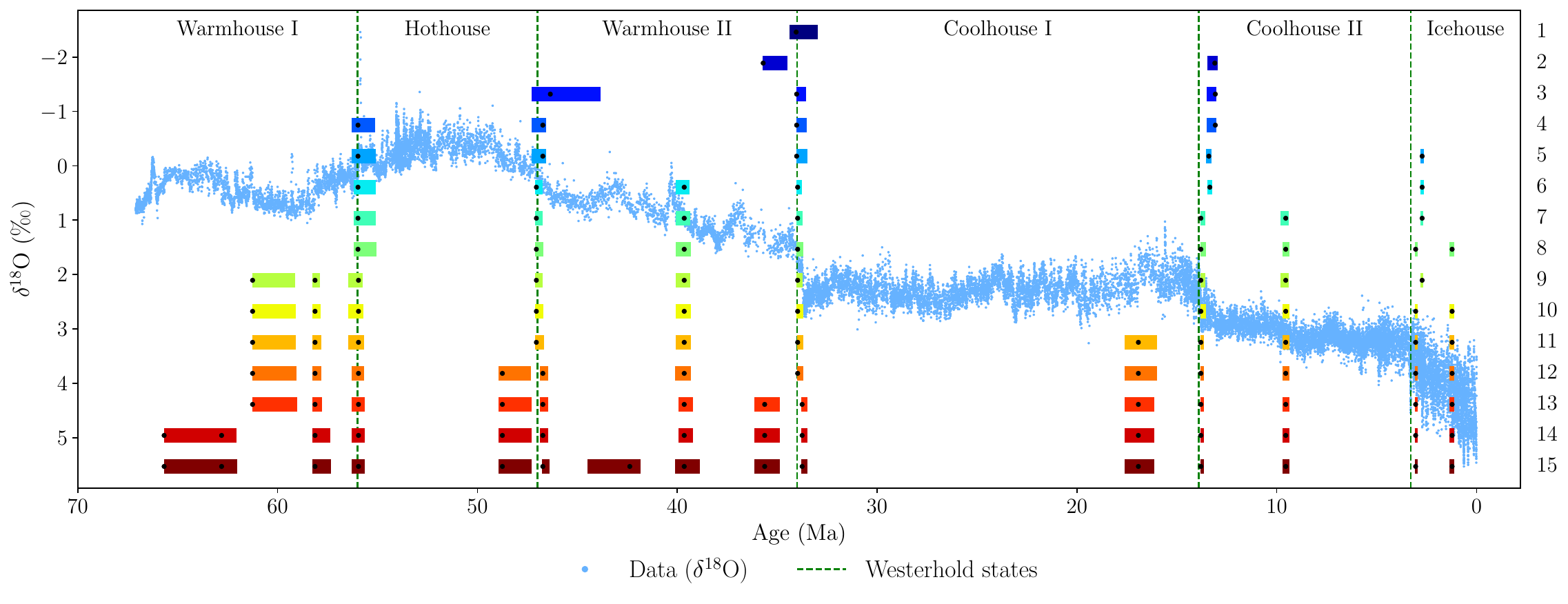 }

    \caption{\fontsize{9}{10}\selectfont
A comparison of estimated breakpoints using the Mean model for one to fifteen breakpoints on 25 kyr binned data. The minimum state length is set to $h=1$ Myr. The black dots represent estimated breakpoints, while colored shaded rectangles indicate 95\% confidence intervals. The results overlay the $\delta^{18}$O data from \cite{Westerhold2020} and their climate states. }
    \label{fig:ManyBPs_mean}
\end{figure}

\subsection{One to fifteen breakpoints: AR model}
\label{1_15_AR}

\begin{figure}[H]
    \centering
        \includegraphics[width=0.8\textwidth]{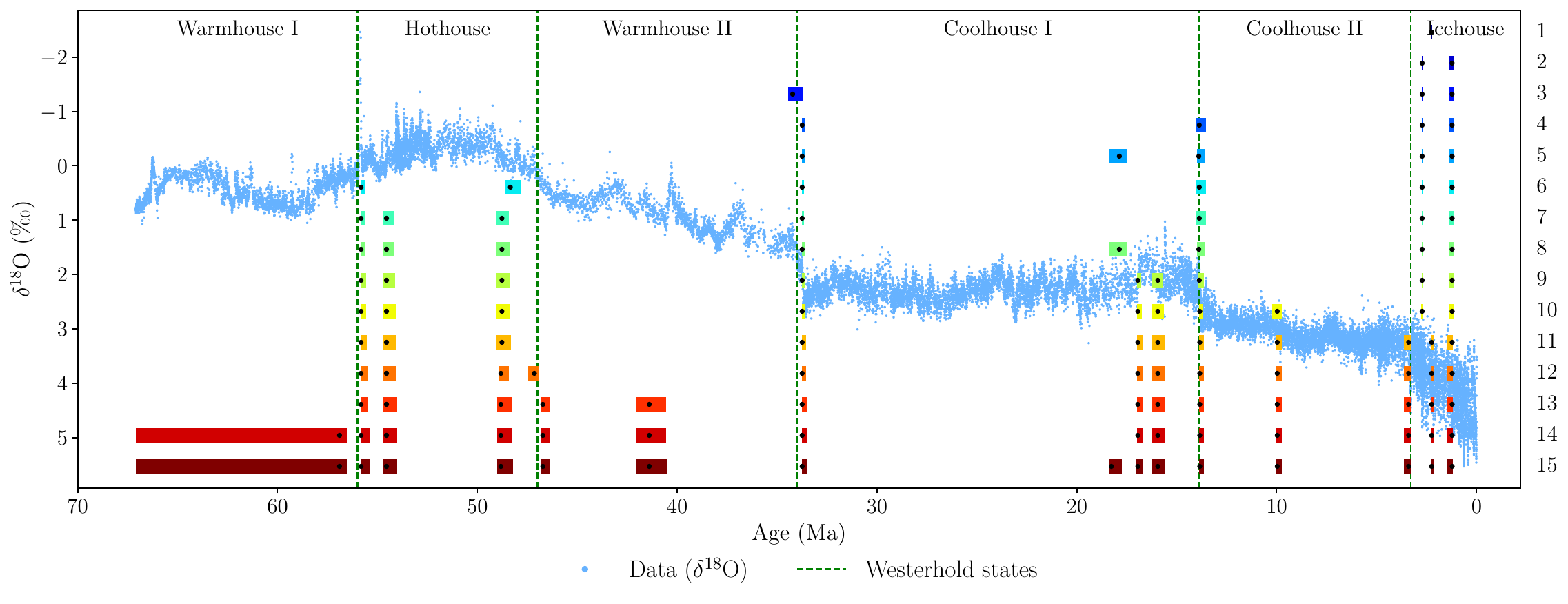}

    \caption{
A comparison of estimated breakpoints using the AR model for one to fifteen breakpoints on 25 kyr binned data. The minimum state length is set to $h=1$ Myr. The black dots represent estimated breakpoints, while colored shaded rectangles indicate 95\% confidence intervals. The results overlay the $\delta^{18}$O data from \cite{Westerhold2020} and their climate states. }
    \label{fig:ManyBPs_AR}
\end{figure}

\section{Tables}

   \vspace{-0.3cm}

\subsection{Summary statistics: State-wise and full sample}
    \label{tab:summary_binned_state}
     \vspace{-0.3cm}

\begin{table}[H]
\fontsize{7}{7}\selectfont 
    \centering
\begin{tabular}{clccccc}
\hline
 Bin size   & State       &   Mean &   Sd. &   Max. &    Min. &   Data points \\
\hline
 5               & Warmhouse I        &  0.417 & 0.249 &  1.07  & -0.215 &          2221 \\
 5               & Hothouse           & -0.269 & 0.261 &  0.391 & -2.014 &          1800 \\
 5               & Warmhouse II       &  0.897 & 0.366 &  1.894 & -0.254 &          2600 \\
 5               & Coolhouse I        &  2.239 & 0.233 &  2.991 &  1.266 &          4020 \\
 5               & Coolhouse II       &  3.072 & 0.237 &  4.172 &  1.885 &          2120 \\
 5               & Icehouse           &  4.037 & 0.463 &  5.405 &  3.05  &           660 \\\hline
 5               & Full sample period &  1.561 & 1.277 &  5.405 & -2.014 &         13421 \\ \hline
 10              & Warmhouse I        &  0.417 & 0.245 &  0.977 & -0.12  &          1111 \\
 10              & Hothouse           & -0.269 & 0.256 &  0.308 & -2.014 &           900 \\
 10              & Warmhouse II       &  0.897 & 0.366 &  1.777 & -0.254 &          1300 \\
 10              & Coolhouse I        &  2.239 & 0.221 &  2.877 &  1.324 &          2010 \\
 10              & Coolhouse II       &  3.072 & 0.228 &  4.122 &  1.975 &          1060 \\
 10              & Icehouse           &  4.034 & 0.447 &  5.33  &  3.181 &           330 \\\hline
 10              & Full sample period &  1.561 & 1.276 &  5.33  & -2.014 &          6711 \\\hline
 25              & Warmhouse I        &  0.418 & 0.237 &  0.912 & -0.065 &           445 \\
 25              & Hothouse           & -0.269 & 0.245 &  0.218 & -1.871 &           360 \\
 25              & Warmhouse II       &  0.898 & 0.358 &  1.688 &  0.01  &           520 \\
 25              & Coolhouse I        &  2.239 & 0.202 &  2.749 &  1.391 &           804 \\
 25              & Coolhouse II       &  3.073 & 0.213 &  3.793 &  2.087 &           424 \\
 25              & Icehouse           &  4.033 & 0.401 &  5.158 &  3.258 &           132 \\\hline
 25              & Full sample period &  1.561 & 1.273 &  5.158 & -1.871 &          2685 \\\hline
 50              & Warmhouse I        &  0.419 & 0.233 &  0.867 & -0.042 &           223 \\
 50              & Hothouse           & -0.268 & 0.233 &  0.197 & -1.871 &           180 \\
 50              & Warmhouse II       &  0.898 & 0.354 &  1.656 &  0.182 &           260 \\
 50              & Coolhouse I        &  2.24  & 0.188 &  2.713 &  1.567 &           402 \\
 50              & Coolhouse II       &  3.072 & 0.206 &  3.72  &  2.156 &           212 \\
 50              & Icehouse           &  4.042 & 0.359 &  4.757 &  3.264 &            66 \\\hline
 50              & Full sample period &  1.562 & 1.271 &  4.757 & -1.871 &          1343 \\\hline
 75              & Warmhouse I        &  0.42  & 0.229 &  0.837 &  0.006 &           148 \\
 75              & Hothouse           & -0.26  & 0.203 &  0.167 & -0.985 &           120 \\
 75              & Warmhouse II       &  0.894 & 0.351 &  1.553 &  0.156 &           173 \\
 75              & Coolhouse I        &  2.239 & 0.181 &  2.717 &  1.691 &           268 \\
 75              & Coolhouse II       &  3.068 & 0.214 &  3.652 &  2.072 &           142 \\
 75              & Icehouse           &  4.041 & 0.351 &  4.753 &  3.283 &            44 \\\hline
 75              & Full sample period &  1.563 & 1.268 &  4.753 & -0.985 &           895 \\\hline
 100             & Warmhouse I        &  0.42  & 0.229 &  0.832 &  0.007 &           112 \\
 100             & Hothouse           & -0.263 & 0.203 &  0.155 & -0.985 &            90 \\
 100             & Warmhouse II       &  0.898 & 0.349 &  1.601 &  0.228 &           130 \\
 100             & Coolhouse I        &  2.241 & 0.175 &  2.685 &  1.739 &           201 \\
 100             & Coolhouse II       &  3.073 & 0.201 &  3.625 &  2.353 &           106 \\
 100             & Icehouse           &  4.047 & 0.344 &  4.673 &  3.4   &            33 \\\hline
 100             & Full sample period &  1.562 & 1.269 &  4.673 & -0.985 &           672 \\\hline
 Without binning & Warmhouse I        &  0.428 & 0.25  &  1.07  & -0.215 &          2761 \\
 Without binning & Hothouse           & -0.279 & 0.255 &  0.391 & -2.46  &          3030 \\
 Without binning & Warmhouse II       &  0.916 & 0.357 &  1.894 & -0.254 &          1786 \\
 Without binning & Coolhouse I        &  2.251 & 0.242 &  3.263 &  1.026 &          6669 \\
 Without binning & Coolhouse II       &  3.102 & 0.254 &  4.49  &  1.84  &          6282 \\
 Without binning & Icehouse           &  4.064 & 0.533 &  5.53  &  2.66  &          3731 \\\hline
 Without binning & Full sample period &  2.128 & 1.445 &  5.53  & -2.46  &         24259 \\
\hline
\end{tabular}
  \vspace{0.1cm}
    \caption{ Summary statistics of the binned data with bin sizes (5, 10, 25, 50, 75, and 100 kyr) and the $\delta^{18}$O data without binning for each of the states identified by \cite{Westerhold2020} and the full sample period.}
\end{table}

\subsection{Estimated breakpoints: 5 breakpoints}
\label{tab:Est_BP_5}

\begin{table}[H]
\fontsize{9pt}{9pt}\selectfont
  \centering
  \begin{tabular}{cccccccc}
    \hline
    Bin size & BP index & \multicolumn{2}{c}{\uline{\quad\quad\quad\quad\quad Mean \quad\quad\quad\;\quad}} & \multicolumn{2}{c}{\uline{\quad\quad\quad\quad Fixed AR \quad\quad\quad\quad}} & \multicolumn{2}{c}{\uline{\quad\quad\quad\quad\quad\quad AR \quad\quad\quad\quad\quad}} \\
    & & Estimate & 95\% CI & Estimate & 95\% CI & Estimate & 95\% CI \\
    \hline
               5 &          1 &          55.965 & (56.085, 55.885) &              55.995 & (56.085, 55.92)   &        33.745 & (33.745, 33.72) \\
               5 &          2 &          46.725 & (46.845, 46.675) &              46.73  & (46.76, 46.68)    &        16.96  & (17.365, 16.78) \\
               5 &          3 &          34.02  & (34.025, 33.915) &              34.05  & (34.075, 34.015)  &        13.825 & (13.84, 13.775) \\
               5 &          4 &          13.36  & (13.395, 13.325) &              13.41  & (13.465, 13.34)   &         9.555 & (9.585, 9.505)  \\
               5 &          5 &           2.735 & (2.845, 2.715)   &               2.74  & (3.1, 2.715)      &         3.36  & (3.815, 3.355)  \\\hline
              10 &          1 &          55.97  & (56.15, 55.79)   &              55.99  & (56.15, 55.88)    &        33.77  & (33.77, 33.72)  \\
              10 &          2 &          46.73  & (46.84, 46.64)   &              46.73  & (46.77, 46.64)    &        17.88  & (18.32, 17.64)  \\
              10 &          3 &          34.02  & (34.03, 33.9)    &              34.15  & (34.18, 34.09)    &        13.82  & (13.84, 13.75)  \\
              10 &          4 &          13.36  & (13.4, 13.3)     &              13.82  & (13.89, 13.72)    &         9.59  & (9.72, 9.45)    \\
              10 &          5 &           2.73  & (2.81, 2.7)      &               2.74  & (3.18, 2.71)      &         2.74  & (2.88, 2.72)    \\\hline
              25 &          1 &          55.975 & (56.3, 55.1)     &              56.025 & (56.575, 55.7)    &        55.825 & (55.85, 55.675) \\
              25 &          2 &          46.725 & (47.3, 46.55)    &              46.725 & (46.825, 46.45)   &        48.35  & (48.625, 47.85) \\
              25 &          3 &          34.025 & (34.05, 33.5)    &              34.15  & (34.225, 34.0)    &        33.75  & (33.75, 33.675) \\
              25 &          4 &          13.4   & (13.525, 13.275) &              13.875 & (13.975, 13.65)   &        13.875 & (14.05, 13.55)  \\
              25 &          5 &           2.725 & (2.8, 2.625)     &               2.775 & (3.075, 2.7)      &         2.575 & (2.6, 2.55)     \\\hline
              50 &          1 &          55.95  & (56.2, 54.6)     &              56     & (57.1, 55.35)     &        56     & (56.65, 55.7)   \\
              50 &          2 &          46.7   & (48.15, 46.45)   &              47.1   & (47.25, 46.55)    &        48.8   & (49.1, 40.45)   \\
              50 &          3 &          34.05  & (34.05, 32.8)    &              34.2   & (34.3, 33.9)      &        33.75  & (33.75, 33.6)   \\
              50 &          4 &          13.8   & (14.15, 13.6)    &              13.85  & (14.0, 13.45)     &        16.95  & (17.35, 16.7)   \\
              50 &          5 &           2.75  & (2.9, 2.5)       &               3.15  & (3.4, 3.0)        &        14.3   & (14.55, 12.8)   \\\hline
              75 &          1 &          55.95  & (56.325, 53.775) &              56.25  & (57.45, 54.75)    &        55.95  & (56.325, 55.5)  \\
              75 &          2 &          46.725 & (50.625, 46.425) &              47.1   & (47.475, 46.425)  &        53.325 & (53.625, 50.1)  \\
              75 &          3 &          34.05  & (34.05, 30.9)    &              34.2   & (34.425, 33.675)  &        34.05  & (34.05, 33.825) \\
              75 &          4 &          13.35  & (13.8, 12.975)   &              13.875 & (14.1, 13.125)    &        16.95  & (17.325, 16.5)  \\
              75 &          5 &           2.775 & (3.375, 2.4)     &               3.15  & (3.525, 2.925)    &        14.475 & (15.075, 14.25) \\\hline
             100 &          1 &          56     & (56.4, 54.0)     &              56.2   & (57.7, 54.5)      &        56     & (56.3, 55.5)    \\
             100 &          2 &          46.7   & (52.5, 46.3)     &              47.1   & (47.7, 46.3)      &        53.4   & (53.8, 52.1)    \\
             100 &          3 &          34.1   & (34.1, 29.4)     &              34.2   & (34.5, 33.4)      &        49.1   & (50.8, 48.8)    \\
             100 &          4 &          13.8   & (14.7, 13.4)     &              13.9   & (14.1, 12.9)      &        34.1   & (34.1, 33.8)    \\
             100 &          5 &           2.9   & (4.2, 2.3)       &               3.4   & (3.8, 3.2)        &        13.8   & (15.7, 12.9)    \\

    \hline
  \end{tabular}
  \label{tab:}
  \vspace{0.2cm}
    \caption{ Estimated breakpoints and their 95\% confidence intervals (in Ma) where the number of breakpoints is fixed to 5, and all values are rounded to three decimals. The table shows estimates for each method across bin sizes 5, 10, 25, 50, 75, and 100 kyr.}
\end{table}

\subsection{Estimated parameters: 5 breakpoints and 25 kyr binned data}

\label{est_params_5}

\begin{table}[H]
\fontsize{9pt}{9pt}\selectfont
  \centering
  \begin{tabular}{lcccccc}
    \hline
     & \multicolumn{2}{c}{\uline{\quad\quad\quad\quad Mean \quad\quad\;\quad}} & \multicolumn{2}{c}{\uline{\quad\quad\quad Fixed AR \quad\quad\quad}} & \multicolumn{2}{c}{\uline{\quad\quad\quad\quad\quad AR \quad\quad\quad\quad}} \\
    Parameter & Estimate & SE & Estimate & SE & Estimate & SE \\ \hline
$c_1$ & 0.418 & 0.051 & 0.069 & 0.008 & -0.001 & 0.026 \\ 
$c_2$ & -0.256 & 0.040 & -0.043 & 0.007 & -0.108 & 0.015 \\ 
$c_3$ & 0.911 & 0.072 & 0.153 & 0.013 & 0.028 & 0.007 \\ 
$c_4$ & 2.247 & 0.017 & 0.373 & 0.031 & 0.660 & 0.061 \\ 
$c_5$ & 3.119 & 0.027 & 0.519 & 0.043 & 0.421 & 0.073 \\ 
$c_6$ & 4.140 & 0.051 & 0.698 & 0.057 & 2.423 & 0.326 \\ 
$\varphi$ & $\times$ & $\times$ & 0.833 & 0.014 & $\times$ & $\times$ \\ 
$\varphi_1$ & $\times$ & $\times$ & $\times$ & $\times$ & 0.990 & 0.054 \\ 
$\varphi_2$ & $\times$ & $\times$ & $\times$ & $\times$ & 0.631 & 0.037 \\ 
$\varphi_3$ & $\times$ & $\times$ & $\times$ & $\times$ & 0.970 & 0.008 \\ 
$\varphi_4$ & $\times$ & $\times$ & $\times$ & $\times$ & 0.706 & 0.027 \\ 
$\varphi_5$ & $\times$ & $\times$ & $\times$ & $\times$ & 0.865 & 0.024 \\ 
$\varphi_6$ & $\times$ & $\times$ & $\times$ & $\times$ & 0.419 & 0.081 \\ 
$\sigma^2_1$ & 0.237 & $\times$ & 0.095 & $\times$ & 0.106 & $\times$ \\ 
$\sigma^2_2$ & 0.255 & $\times$ & 0.154 & $\times$ & 0.140 & $\times$ \\ 
$\sigma^2_3$ & 0.347 & $\times$ & 0.112 & $\times$ & 0.107 & $\times$ \\ 
$\sigma^2_4$ & 0.210 & $\times$ & 0.141 & $\times$ & 0.140 & $\times$ \\ 
$\sigma^2_5$ & 0.208 & $\times$ & 0.111 & $\times$ & 0.116 & $\times$ \\ 
$\sigma^2_6$ & 0.351 & $\times$ & 0.340 & $\times$ & 0.315 & $\times$ \\ 

    \hline
  \end{tabular}
  \label{}
  \vspace{0.2cm}
               \caption{ Estimated parameters and their corresponding standard errors (SE) for each model specification. Parameters absent in a given model specification are denoted by $\times$. The number of breakpoints is set to 5, and the parameters are estimated with a binning frequency of 25 kyr and $h=2.5$ Myr. All values are rounded to three decimals.}
\end{table}